\documentclass[aps,prd,preprint,groupedaddress]{revtex4}

\usepackage{epsfig}

\begin{document}

\def\d{{\rm d}}
\def\F{\tilde{F}}
\def\p{I\!\!P}
\def\R{\tilde{R}}

\def\lp{\left. }
\def\rp{\right. }
\def\lr{\left( }
\def\rr{\right) }
\def\le{\left[ }
\def\re{\right] }
\def\lg{\left\{ }
\def\rg{\right\} }
\def\lb{\left| }
\def\rb{\right| }

\def\beq{\begin{equation}}
\def\eeq{\end{equation}}
\def\bea{\begin{eqnarray}}
\def\eea{\end{eqnarray}}

\preprint{DESY 09-130}
\preprint{LPSC 09-112}
\title{Survival probability for diffractive dijet production in
       $p\bar{p}$  collisions from next-to-leading order calculations}
\author{Michael Klasen}
\email[]{klasen@lpsc.in2p3.fr}
\affiliation{Laboratoire de Physique Subatomique et de Cosmologie,
 Universit\'e Joseph Fourier/CNRS-IN2P3/INPG, 53 Avenue des Martyrs,
 F-38026 Grenoble, France}
\author{Gustav Kramer}
\affiliation{{II.} Institut f\"ur Theoretische Physik, Universit\"at
 Hamburg, Luruper Chaussee 149, D-22761 Hamburg, Germany}
\date{\today}
\begin{abstract}
 We perform next-to-leading order calculations of the single-diffractive and
 non-diffractive cross sections for dijet production in proton-antiproton
 collisions at the Tevatron. By comparing their ratio to the data published
 by the CDF collaboration for two different center-of-mass energies, we deduce
 the rapidity-gap survival probability as a function of the momentum fraction
 of the parton in the antiproton. Assuming Regge factorization, this probability
 can be interpreted as a suppression factor for the diffractive structure function
 measured in deep-inelastic scattering at HERA. In contrast to the observations
 for photoproduction, the suppression factor in proton-antiproton collisions
 depends on the momentum fraction of the parton in the Pomeron even at
 next-to-leading order.
\end{abstract}
\maketitle


\vspace*{-100mm}
\noindent DESY 09-130\\
\noindent LPSC 09-112\\
\vspace*{77mm}


\section{Introduction}

Diffractive events in high-energy $p\bar{p}$ or $ep$ collisions are
characterized by the presence of a leading proton or antiproton, which
remains intact, and/or by a rapidity gap, defined as a (pseudo-)rapidity
region devoid of particles. Theoretically, diffractive interactions are
described in the framework of Regge theory \cite{1} as the exchange of a
trajectory with vacuum quantum numbers, the so-called Pomeron ($\p$)
trajectory. Diffractive scattering involving hard processes (hard
diffraction) such as the production of high-$E_T$ jets has been studied
experimentally to investigate the parton content of the Pomeron (or
additional lower-lying Regge poles). In this framework, $p\bar{p}$ hard
diffraction can be expressed as a two-step process, $p+\bar{p} \to p+
\p+\bar{p}' \to 2~{\rm jets}+\bar{p}'+X$, and similarly diffractive
deep-inelastic scattering (DDIS) as $\gamma^{*}+p \to \gamma^{*}+\p+p'
\to p'+X$. The subprocess $\gamma^{*}+\p \to X$ is interpreted as
deep-inelastic scattering (DIS) on the Pomeron target for the case that
the virtuality of the exchanged photon $Q^2$ is sufficiently large. In
analogy to DIS on a proton target, $\gamma^{*}+p \to X$, the cross
section for DIS on a Pomeron target is expressed as a convolution of
partonic cross section and universal parton distribution functions
(PDFs) of the Pomeron. The partonic cross sections are the same as for
$\gamma^{*}p$ DIS. The Pomeron PDFs are multiplied with vertex functions
for the vertex $p \to \p+p'$, yielding the diffractive PDFs (DPDFs). The
additional vertex functions depend on the fractional momentum loss $\xi$
and the four-momentum transfer squared $t$ of the recoiling proton. The
DDIS experiments measure the diffractive structure function of the
proton $F_2^{\rm D}(\xi,\beta,Q^2)$ integrated over $t$, where $\beta=x/\xi$
is the momentum fraction of the parton in the Pomeron and $Q^2$ is the
virtuality of the $\gamma^{*}$. The $Q^2$ evolution of the DPDFs is
calculated with the usual DGLAP \cite{2} evolution equations known from
$\gamma^{*}+p \to X$ DIS. Except for the $Q^2$ evolution, the DPDFs are
not calculable in the framework of perturbative QCD and must be
determined from experiment. Such DPDFs have been obtained from the HERA
inclusive measurements of $F_2^{\rm D}$ \cite{3,4}. The presence of a hard
scale such as the squared photon virtuality $Q^2$ in DIS or a large
transverse jet energy $E_T^{\rm jet}$ in hard diffractive processes, as
for example in $p\p \to {\rm jets}+X$ or $\gamma \p \to {\rm jets}+X$,
allows for the calculation of the corresponding partonic cross sections
using perturbative QCD. The central issue is whether such hard
diffractive processes obey QCD factorization, i.e.\ can be calculated
in terms of parton-level cross sections convolved with universal DPDFs.

For DIS processes, QCD factorization has been proven to hold \cite{5},
and DPDFs have been extracted at low and intermediate $Q^2$ \cite{3,4}
from high-precision inclusive measurements of the process $e+p \to e'+
p'+X$ using the usual DGLAP evolution equations. The proof of the
factorization formula also appears to be valid for the direct part of
photoproduction ($Q^2 \simeq 0$) of jets \cite{5}. However,
factorization does not hold for hard processes in diffractive
hadron-hadron scattering. The problem is that soft interactions
between the ingoing hadrons and their remnants occur in both the
initial and final states. This was also the result of experimental
measurements by the CDF collaboration at the Tevatron \cite{6}, where
it was found that the single-diffractive dijet production cross
section was suppressed by up to an order of magnitude as compared to
the prediction based on DPDFs determined earlier by the H1
collaboration at HERA \cite{7}. In the CDF experiment \cite{6}, the
suppression factor was determined by comparing single-diffractive (SD)
and non-diffractive (ND) events. SD events are triggered on a leading
antiproton in the Roman pot spectrometer and at least one jet, while
the ND trigger requires only a jet in the CDF calorimeters. The ratio
$R(x,\xi,t)$ of SD to ND dijet production rates $N_{\rm JJ}$ is in a
first approximation proportional to the ratio of the corresponding
structure functions $F_{\rm JJ}$, i.e.\
\beq
 R(x,\xi,t) = \frac{N_{\rm JJ}^{\rm SD}(x,Q^2,\xi,t)}{N_{\rm JJ}^{\rm ND}(x,Q^2)}
 \approx \frac{F_{\rm JJ}^{\rm SD}(x,Q^2,\xi,t)}{F_{\rm JJ}^{\rm ND}(x,Q^2)}.
\eeq
An approximation to the SD structure function $F_{\rm JJ}^{\rm SD}(x,Q^2,\xi,t)$,
$\F_{\rm JJ}^{\rm D}(\beta)$, was obtained by multiplying the above ratio of
rates by the known effective
\beq
  F_{\rm JJ}^{\rm ND}(x) = x[g(x)+\frac{4}{9}\sum_{i}q_i(x)]
\eeq
after integrating this ratio over $\xi$ and $t$ and changing variables
from $x$ to $\beta$ using $x \to \beta \xi$. The result
was then compared to the DPDFs from H1 \cite{7} using the same
approximate formula, Eq.\ (2), relating the structure function to
gluon and quark DPDFs as in the ND case. The above formula for the
ratio $R(x,\xi,t)$ is certainly not sufficient for estimating the
suppression factor for diffractive dijet production in $p\bar{p}$
collisions. It is based on a leading order (LO) calculation of the
cross section in the numerator and in the denominator. Furthermore,
it is assumed that the convolutions of the PDFs in the numerator
and the denominator with the partonic cross sections are identical
and drop out in the ratio together with the PDFs for the ingoing
proton. These approximations are not valid in next-to-leading order
(NLO), where, in particular, the cross sections in the numerator
and denominator depend on the jet algorithm and
on the kinematics of the SD and ND processes.

Since 2002, the two HERA collaborations have presented results for
diffractive dijet photoproduction in order to establish a possible
suppression factor. The factorization breaking was first
investigated on the basis of NLO predictions by us in 2004
\cite{8,9} by comparing to preliminary H1 data \cite{10}. Already
in 2004 it became clear that in photoproduction the breaking could
be shown only by comparing with NLO predictions, which produced by
a factor of two larger cross sections than the LO predictions.
Concerning factorization breaking, the conclusions were the same
based on a preliminary ZEUS analysis \cite{11}. Both collaborations, H1
and ZEUS, have now published their final experimental data \cite{12,13}.
Whereas H1 confirm in \cite{12} their earlier findings based on the
analysis of their preliminary data and preliminary DPDFs, the ZEUS
collaboration \cite{13} reached somewhat different conclusions from their
analysis. In particular, the H1 collaboration \cite{12} obtained a global
suppression of their measured cross sections as compared to the NLO
calculation of approximately $S = 0.5$. In addition they concluded that
also the direct cross section together with the resolved one does not
obey factorization. The ZEUS collaboration, however, concluded from
their analysis \cite{13} that, within the large uncertainties of the NLO
calculations and the differences in the DPDF input, their data are
compatible with the NLO QCD calculation, i.e.\ a suppression could not
be deduced from their data. In several recent reviews, we have shown,
however, that the ZEUS data are compatible with the older H1 \cite{12}
and with even more recent H1 data \cite{14}, if one adjusts the ZEUS
large rapidity-gap inclusive DIS diffractive data to the analogous H1
data, which are the basis of the recent H1 DPDFs \cite{4} and which are
used to predict the diffractive dijet photoproduction cross sections.
In these recent reviews \cite{15} we also investigated whether the NLO
prediction with resolved suppression only, which would be more in line
with the findings in \cite{5}, will also describe the H1 and ZEUS data
in a satisfactory way. The result is, that this is indeed possible,
and the resolved suppression factor is of the order of $S \approx 0.3$. For the
global suppression, i.e.\ direct and resolved component equally, the
suppression factor is larger, and in addition, depends on $E_T^{\rm
jet}$, which is not the case for the resolved suppression only.

In this work we want to bring the theoretical analysis of diffractive
dijet production in $p\bar{p}$ collisions to the same level as has been
done for diffractive dijet photoproduction, i.e.\ to calculate the cross
sections up to NLO and then compare with the CDF data, to establish the
suppression factor in the Tevatron energy range. For this purpose we
shall calculate the ratio $R(x,\xi,t)$. For this we need the NLO cross
sections for SD and ND with the cuts as in the CDF measurements.
The outline of the paper is as follows. In Section 2 we shall describe
shortly the kinematic restrictions for the CDF analysis based on
measurements at Run I for $\sqrt{s}=1800$ GeV \cite{6} and on
measurements at $\sqrt{s}=630$ GeV and $\sqrt{s}=1800$ GeV \cite{16}
obtained for comparison at two different center-of-mass
energies. In this section we shall also specify the various inputs for
our calculation. Our results and the comparison with the CDF data are
presented in Sect.\ 3. The first $1800$ GeV data are compared with the
calculations in Sect.\ 3.1. The comparative study of the 630 and the new
1800 GeV cross sections are presented in Sect.\ 3.2. An interpretation
of the observed suppression factor is given in Sect.\ 3.3.
Sect.\ 4 contains a summary and our conclusions.


\section{Kinematic cuts and input for the calculations}

The data, which we want to compare our NLO calculations with, are
published in Ref.\ \cite{6} and Ref.\ \cite{16}. In the first paper
\cite{6}, the CDF collaboration measured non-diffractive and
single-diffractive dijet cross sections at a center-of-mass energy of
$\sqrt{s}=1800$ GeV using Run IC (1995-1996) data. From an inclusive
sample of single-diffraction (SD) events, $\bar{p}p \to \bar{p'}X$,
triggering on a $\bar{p}$ detected in a forward Roman pot spectrometer,
a diffractive dijet subsample with transverse energy $E_T^{\rm jet}>7$
GeV was selected. In addition to the two leading jets, these events
contain other lower-$E_T$ jets. Similarly, a non-diffractive (ND) dijet
sample was selected. From the $E_T$ and the rapidity $\eta$ of the jets,
the fraction $x_{\bar{p}}$ of the momentum of the antiproton carried by
the struck parton was calculated, where $x_{\bar{p}}$ is given by
\beq
    x_{\bar{p}}=\frac{1}{\sqrt{s}} \sum_{i} E_T^i e^{-\eta^i}.
\eeq
The jets were detected and their energy measured by calorimeters
covering the pseudorapidity range $|\eta| < 4.2$. The $E_T^{\rm jet}$
was defined as the sum of the calorimeter $E_T$'s within an $\eta-\phi$
cone of radius 0.7. The jet energy correction included a subtraction of
an average underlying event of $E_T$ of 0.54 (1.16) GeV for diffractive
(non-diffractive) events. The recoil antiproton fractional momentum loss
$\xi$ and four-momentum transfer squared $t$ were in the range $0.035<\xi
<0.095$ and $(-t) < 3$ GeV$^2$, respectively, which was in the final
sample restricted to $(-t) < 1$ GeV$^2$. In the second paper \cite{16},
the study of diffractive dijet events was extended to $\sqrt{s}=630$
GeV. These data were compared to new measurements at $\sqrt{s}=1800$ GeV
in order to test Regge factorization. This study is similar to the
previous diffractive dijet study in experimental setup and methodology.
For the SD sample, the $\xi$-region is the same, $0.035<\xi<0.095$, but
$(-t)<0.2$ GeV$^2$. Again in the SD sample events with at least two
jets with $E_T^{\rm jet}>7$ GeV were selected, where again the
$E_T^{\rm jet}$ was defined as the sum of the calorimeter $E_T$'s within
a cone of 0.7 in $\eta-\phi$ space. The jet energy correction included
a subtraction of an average underlying event of 0.5 (0.9) GeV for SD (ND)
events.

The calculation of the cross sections for dijet production in
non-diffractive and single-diffractive processes has been performed up
to NLO. For the comparison we have calculated these cross sections also
in LO. For our calculations, we rely on our work on dijet production in
the reaction $\gamma+p \to {\rm jets}+X$ \cite{17}, in which we have
calculated the cross sections for inclusive one-jet and two-jet
production up to NLO for both the direct and the resolved contribution.
The version for the resolved contribution can be used immediately for
two-jet production in $p\bar{p}$ collisions by substituting for the
photon PDF the antiproton PDF (for ND) or the Pomeron PDF (for SD). For
the (anti-)proton PDF we have chosen the version CTEQ6.6M \cite{18} for
the NLO calculation with $N_f=5$ active flavors. The strong coupling
constant $\alpha_s$ is calculated from the two-loop formula with
$\Lambda^{(5)}_{\overline{\rm MS}}=226$ MeV. For the calculation in LO we
have chosen CTEQ6L1 \cite{19} with $\alpha_s$ determined from the
one-loop formula and $\Lambda^{(5)}=165$ MeV. The diffractive PDFs are
taken from the recent H1 fits to the inclusive diffractive DIS data
\cite{4}. They are only available at NLO and come in two versions,
'H1 2006 fit A' and 'H1 2006 fit
B'. These differ mostly in the gluon density, which is poorly
constrained by the inclusive diffractive scattering data, since there
is no direct coupling of the photon to gluons, so that the gluon density
is constrained only through the evolution. The 'H1 2006 fit A' has a
much larger gluon for larger momentum fractions $\beta$ at the starting
scale of $Q_0=\sqrt{8.5}$ GeV than 'fit B', which leads to a larger
gluon also for larger scales $Q$. The original fit on the data in
\cite{4} is performed with $N_f=3$ massless flavors. The production of
charm quarks was treated in the Fixed-Flavor Number Scheme (FFNS) in NLO
with non-zero charm-quark mass yielding a diffractive $F_2^c$. This
$F_2^c$ is contained in the 'H1 fit 2006 A, B' parameterizations and is
then converted into a charm PDF. The H1 collaboration constructed a
third set of DPDFs, which is called the 'H1 2007 fit jets' and which is
obtained through a simultaneous fit to the diffractive DIS inclusive and
dijet cross sections \cite{20}. It is performed under the assumption
that there is no factorization breaking in the diffractive dijet cross
sections. Including the diffractive DIS dijet cross section in the
analysis leads to additional constraints, mostly on the diffractive
gluon distribution. On average, the 'H1 2007 fit jets' is similar to the
'H1 2006 fit B', except for the gluon distribution at large momentum
fraction and small factorization scale. The DPDFs of H1 contain as a
factor the vertex function $f_{\p/p}(\xi,t)$, which describes the
coupling of the Pomeron to the proton, i.e.\ the proton-proton-Pomeron
vertex. This vertex function is parameterized by the Pomeron trajectory
$\alpha_{\p}(t)$ and an additional exponential dependence on $t$. This
function is used also for our calculations, as it has been determined by
the H1 collaboration when fitting their data. The normalization factor
$N$ of this function is included in the Pomeron PDFs. Therefore the H1
DPDFs are products of the Pomeron flux factors and the Pomeron PDFs.
These H1 DPDFs include also low-mass proton dissociative processes with
invariant mass $M_Y<1.6$ GeV, which increases the inclusive diffractive
DIS cross section as compared to cross sections with a pure
(anti-)proton final state. We have to keep this in mind, when we compare
to the CDF data, which use a forward Roman pot spectrometer to trigger
on the final antiproton and therefore have no antiproton dissociative
contributions.


\section{Results}

In this section we present our results and compare them to the
experimental data obtained with $\sqrt{s}=1.8$ TeV in \cite{6} and to
the more recent data with $\sqrt{s}=0.63$ TeV and $\sqrt{s}=1.8$ TeV
published in \cite{16}. In this latter publication, the kinematic
constraints differ in some points from the constraints used in \cite{6}.
First we compare to the normalized differential cross sections $\d\sigma
/\d \overline{E_T}$ and $\d\sigma/\d\overline{\eta}$ for non-diffractive and diffractive dijet
production. Second, the ratio $\R(x_{\bar{p}})$ of the number of SD dijet
events to
the number of ND dijets is compared to the CDF data. This function
$\R(x_{\bar{p}})$, obtained by integrating the cross sections in the numerator of
Eq.\ (1) over
$\xi$ and $t$, is the main result, and from the theoretical and
experimental distribution as a function of $x_{\bar{p}}$ the suppression
factor $\R^{\rm exp}(x_{\bar{p}})/\R^{\rm (N)LO}(x_{\bar{p}})$ can be
deduced and
can be studied for the three H1 DPDFs, 'H1 2006 fit A', 'H1 2006 fit B'
\cite{4} and 'H1 2007 fit jets' \cite{20} for the NLO and the LO (which
has been done only for the 'H1 2006 fit B') calculations.


\subsection{Comparison with 1800 GeV data}

First we have calculated the distribution $\frac{1}{\sigma}\frac{\d
\sigma}{\d\overline{E_T}}$ as a function of $\overline{E_T}=(E_T^{\rm
jet1}+E_T^{\rm jet2})/2$, where $E_T^{\rm jet1} (E_T^{\rm jet2})$
refers to the jet with the largest (second largest) $E_T$ for ND and SD
dijet production with $\sqrt{s}=1800$ GeV center-of-mass energy,
integrated over the rapidities of the jets in the range $|\eta|<4.2$.
Jets are defined with the usual cone algorithm within a chosen $\eta-
\phi$ cone of radius $R=0.7$ and a partonic distance $R_{\rm sep}=1.3R$
to match the experimental analysis \cite{20a}.
$\sigma$ is the integrated cross
section with the cut $E_T^{\rm jet1(2)} > 7.0(6.5)$ GeV. The lower
limit of the leading and subleading jet differ slightly in order to
avoid infrared sensitivity in the computation of the NLO cross sections,
when integrated over $\overline{E_T}$ \cite{21}. Unfortunately in the
experimental analysis such an asymmetric choice of $E_T^{\rm jet1}$ and
$E_T^{\rm jet2}$ has not been made, since both $E_T^{\rm jet1}$ and
$E_T^{\rm jet2}$ are restricted by $E_T^{\rm jet1,2} > 7.0$ GeV, so
that we do not know whether the choice $E_T^{\rm jet2} > 6.5$ GeV is in
accord with the experimental analysis. Therefore we have also varied the
$E_T^{\rm jet2}$ cut slightly to $E_T^{\rm jet2} > 6.6$ GeV. The
results for $\frac{1}{\sigma}\frac{\d\sigma}{\d\overline{E_T}}$ are
%
\begin{figure}
 \centering
 \includegraphics[width=0.49\columnwidth]{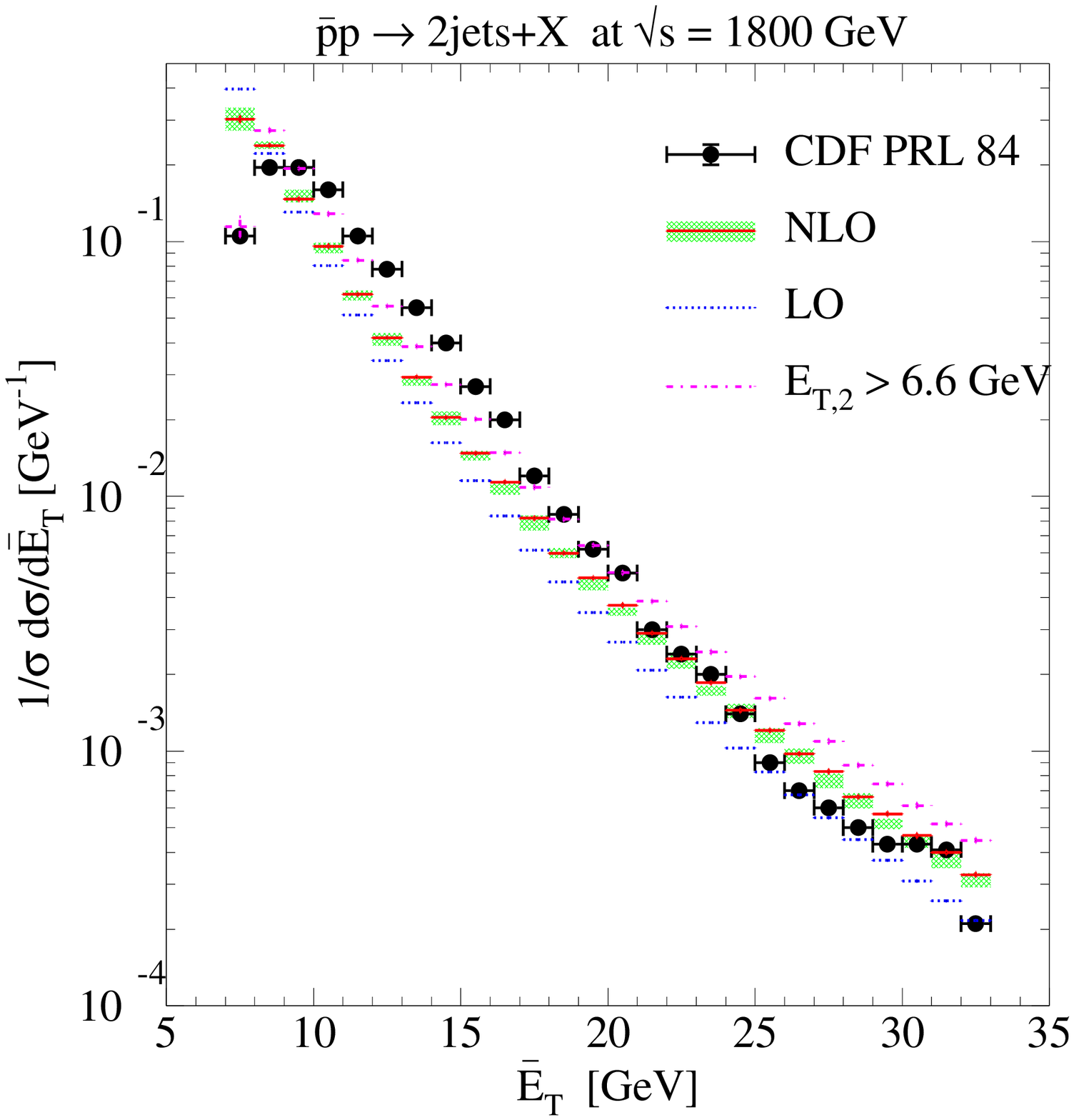}
 \includegraphics[width=0.49\columnwidth]{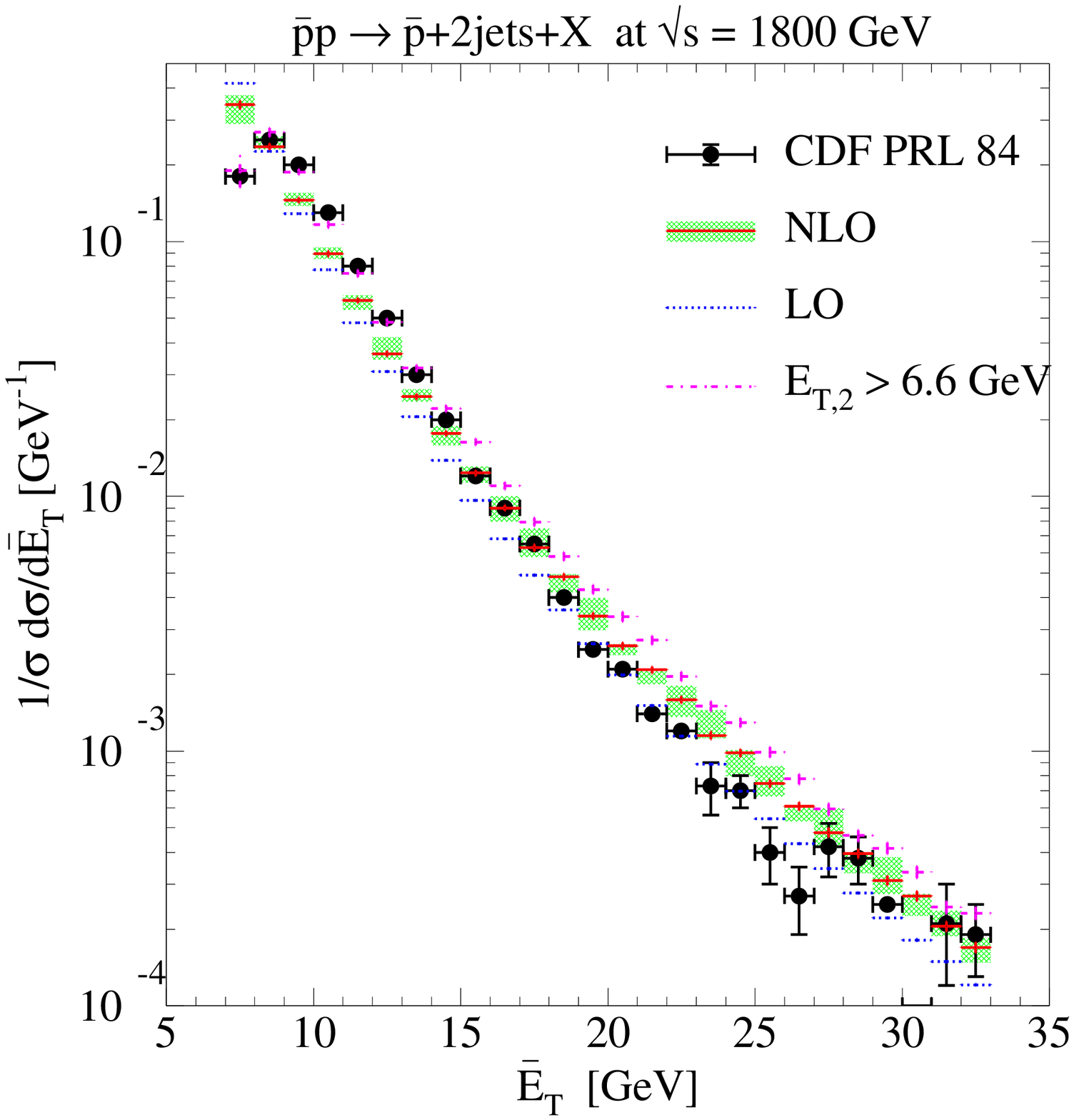}
 \caption{\label{fig:1}Normalized average transverse-energy
 distributions of the non-diffractive (left) and single-diffractive
 (right) dijet cross section at Run I of the Tevatron. The CDF data
 (points) are compared with our predictions at NLO (full) and LO
 (dotted) and also with a varied cut on the subleading jet $E_T$
 (dot-dashed). The NLO scale uncertainty is shown as a shaded band
 (color online).}
\end{figure}
%
shown in Fig.\ 1 (left) for $E_T^{\rm jet2} > 6.5$ GeV (full histogram)
and $E_T^{\rm jet2} > 6.6$ GeV (dot-dashed histogram), respectively.
Together with the NLO cross section, we also show the LO cross section
(dotted histogram) and the scale variation of the NLO result (shaded
band), where the renormalization and factorization scales are varied
simultaneously by factors of 0.5 and 2.0 compared to the default scale,
which is chosen equal to $E_T^{\rm jet1}$, i.e.\ the largest $E_T$ of
both jets. As is often the case, the scale uncertainty is relatively
small in the normalized distributions.
In Fig.\ 1 (left) we have included also the measured cross
section from \cite{6}, which unfortunately is given without the
experimental uncertainties. Besides the statistical errors,
which should be similar to those in the single-diffractive distributions
given the similar number of ND and SD events \cite{20a}, there are
also systematic errors, as for example those associated with the jet
energy scale. These would be needed for a fair comparison. Second, the
theoretical cross sections should be corrected for hadronization
effects when comparing to data. These are not known to us, but could be
calculated through Monte Carlo models. If we compare the calculations
in Fig.\ 1 (left) with the data, we observe that the results with
$E_T^{\rm jet2} > 6.5$ GeV agree reasonably well with the data in the
large $\overline{E_T}$ range, $\overline{E_T} > 20$ GeV, but much less
for the low and medium $\overline{E_T}$ range. Conversely, for
$E_T^{\rm jet2} > 6.6$ GeV the small and medium $\overline{E_T}$ range
agrees better and the large $\overline{E_T}$ range less. The
experimental errors will be larger in the large $\overline{E_T}$ range.
Therefore the cross section with the cut $E_T^{\rm jet2} > 6.6$ GeV
would be preferred, in particular also because we have perfect
agreement in the first, second and third $\overline{E_T}$ bin, which
are the most important ones for the integrated cross section $\sigma$,
which determines the normalization.

The equivalent comparison for the SD dijet $\overline{E_T}$-distribution
is shown in Fig.\ 1 (right) for $E_T^{\rm jet1(2)}>7.0
(6.5)$ GeV (full) and $E_T^{\rm jet1(2)}>7.0(6.6)$ GeV (dot-dashed).
Here we have chosen only the 'H1 2006 fit B' as DPDF. The comparison
of data versus theory in Fig.\ 1 (right) shows the same pattern as for
the ND distributions in Fig.\ 1 (left). In general the agreement with
the data is even somewhat better now, in particular for the $E_T^{\rm
jet1(2)}>7.0(6.6)$ GeV cut. As for the ND distribution, we present also the LO
prediction (dotted). From the unnormalized distributions (not shown),
we obtain ratios of NLO to the LO cross sections ($K$-factors),
which increase from relatively small values of 0.5 (0.6) in the
infrared-sensitive region close to the $E_T^{\rm jet1,2}$ cuts 
to unity at larger $\overline{E_T}$ for the ND (SD) cross sections,
indicating good perturbative stability and no sensitivity to the
cut on $E_T^{\rm jet2}$ there.

The ND $\bar{\eta}$-distribution $\frac{1}{\sigma}\frac{\d\sigma}
{\d\bar{\eta}}$, where $\bar{\eta}=(\eta^{\rm jet1}+\eta^{\rm
jet2})/2$, is plotted in Fig.\ 2 (left), again for the two
%
\begin{figure}
 \centering
 \includegraphics[width=0.49\columnwidth]{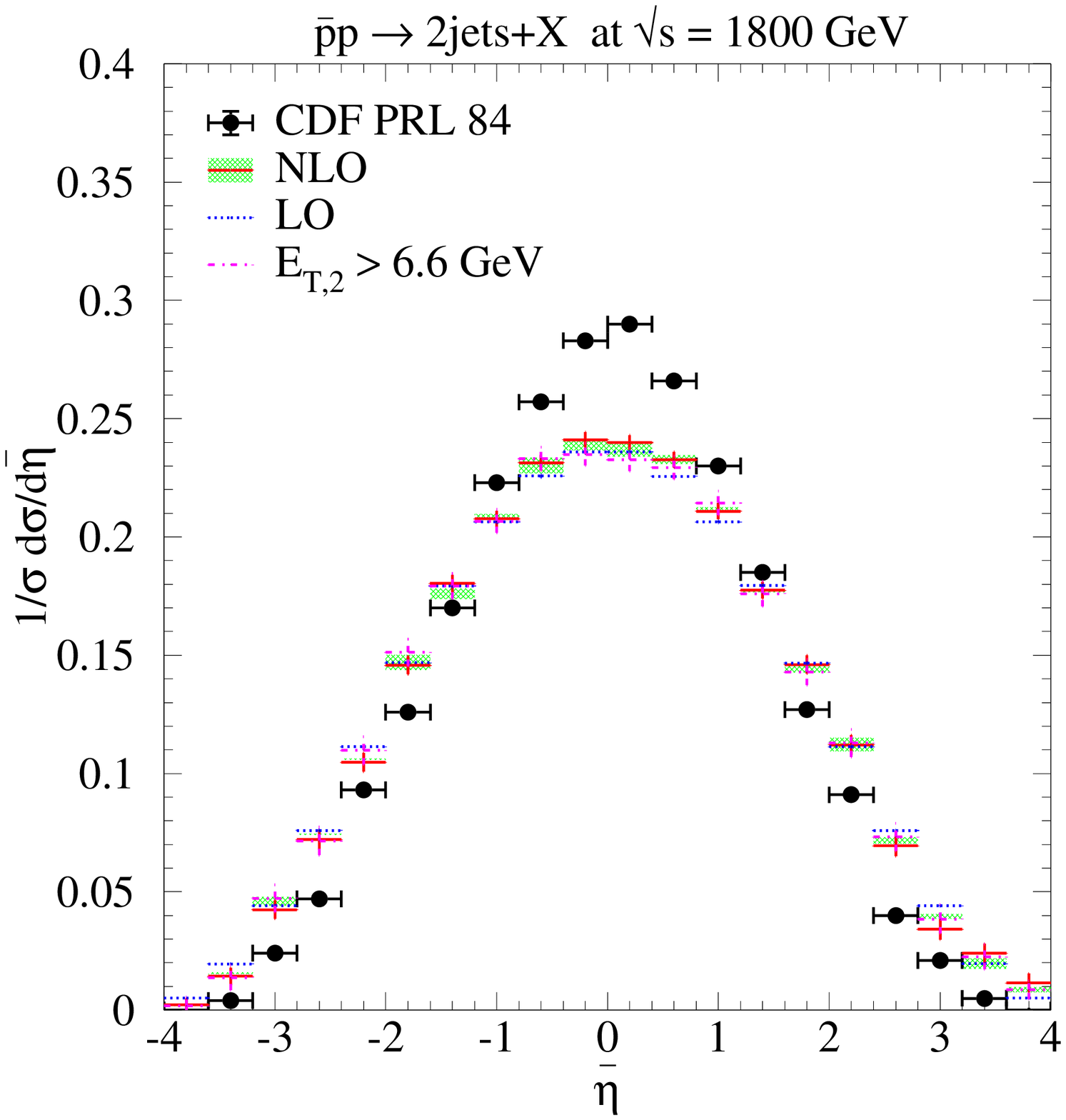}
 \includegraphics[width=0.49\columnwidth]{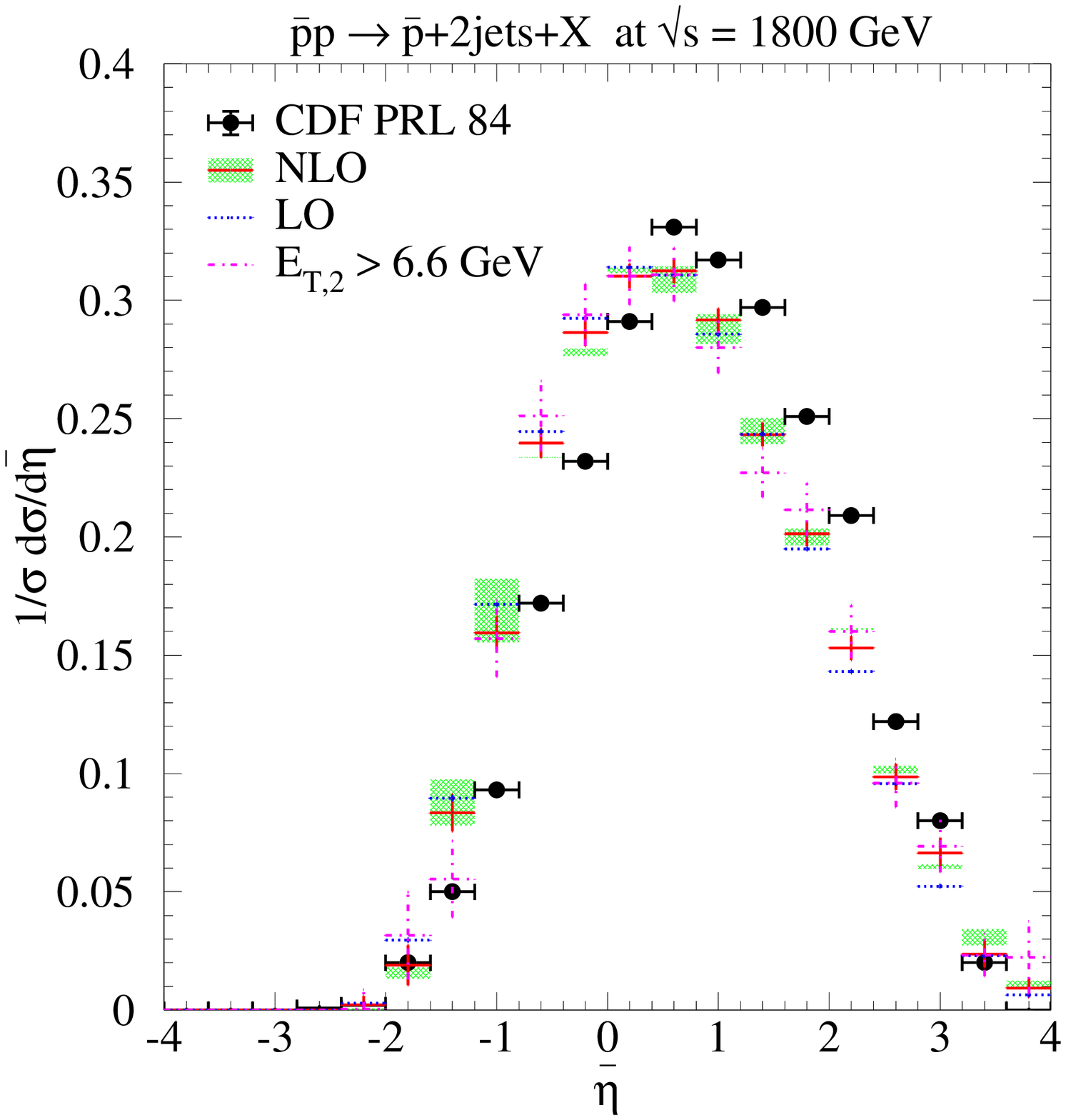}
 \caption{\label{fig:2}Same as Fig.\ 1, but for the normalized average
 rapidity distributions (color online).}
\end{figure}
%
$E_T^{\rm jet2}$ cuts, and the SD $\bar{\eta}$-distribution in Fig.\ 2
(right). For ND and SD, the two choices for the $E_T^{\rm jet2}$ have
little influence on the result. The cross sections are somewhat smaller
(by about 20\%) at the maximum as compared to the ND experimental data,
again given without experimental errors, but are in agreement with the SD
data. The theoretical diffractive $\bar{\eta}$-distribution is boosted
towards positive $\bar{\eta}$, as is the experimental one. We conclude
from these comparisons that there is satisfactory agreement between the
measured $\overline{E_T}$ and $\overline{\eta}$ distributions and our
theoretical predictions based on the 'H1 2006 fit B' DPDF. This
motivates us to move on to the calculation of the ratio
$\R(x_{\bar{p}})$ of SD to ND cross sections. Actually, if the
experimental cross sections above had been known to us in the
unnormalized form, we would have been in the position to deduce
suppression factors as a function of $\overline{E_T}$ and $\bar{\eta}$.

The ratio $\R(x_{\bar{p}})$ of the SD to ND cross sections is evaluated
as a function of $x_{\bar{p}}$, the fraction of the momentum of the
antiproton transferred to the struck parton. It is calculated from the
$E_T^i$ and $\eta^i$ of the jets with the relation in Eq.\ (3), where
the sum over $i$ is taken over the two leading jets plus the next highest
$E_T$ jet with $E_T > 5$ GeV. The cross section in the numerator is
integrated over $\xi$ in the range $0.035 < \xi < 0.095$ and over
$(-t)$ in the range $(-t) < 1$ GeV$^2$ and over the $E_T^{\rm jet}$ of
the highest and second highest $E_T^{\rm jet}$ with $E_T^{\rm jet1(2)} >
7.0(6.5)$ GeV for both the SD and ND jet sample. As already mentioned,
the data samples have the constraint $E_T^{\rm jet1(2)} > 7.0(7.0)$
GeV. We have checked that the choice $E_T^{\rm jet2} > 6.6$ GeV for the ND
and SD cross sections has negligible influence on $\R(x_{\bar{p}})$.
%
\begin{figure}
 \centering
 \includegraphics[width=0.49\columnwidth]{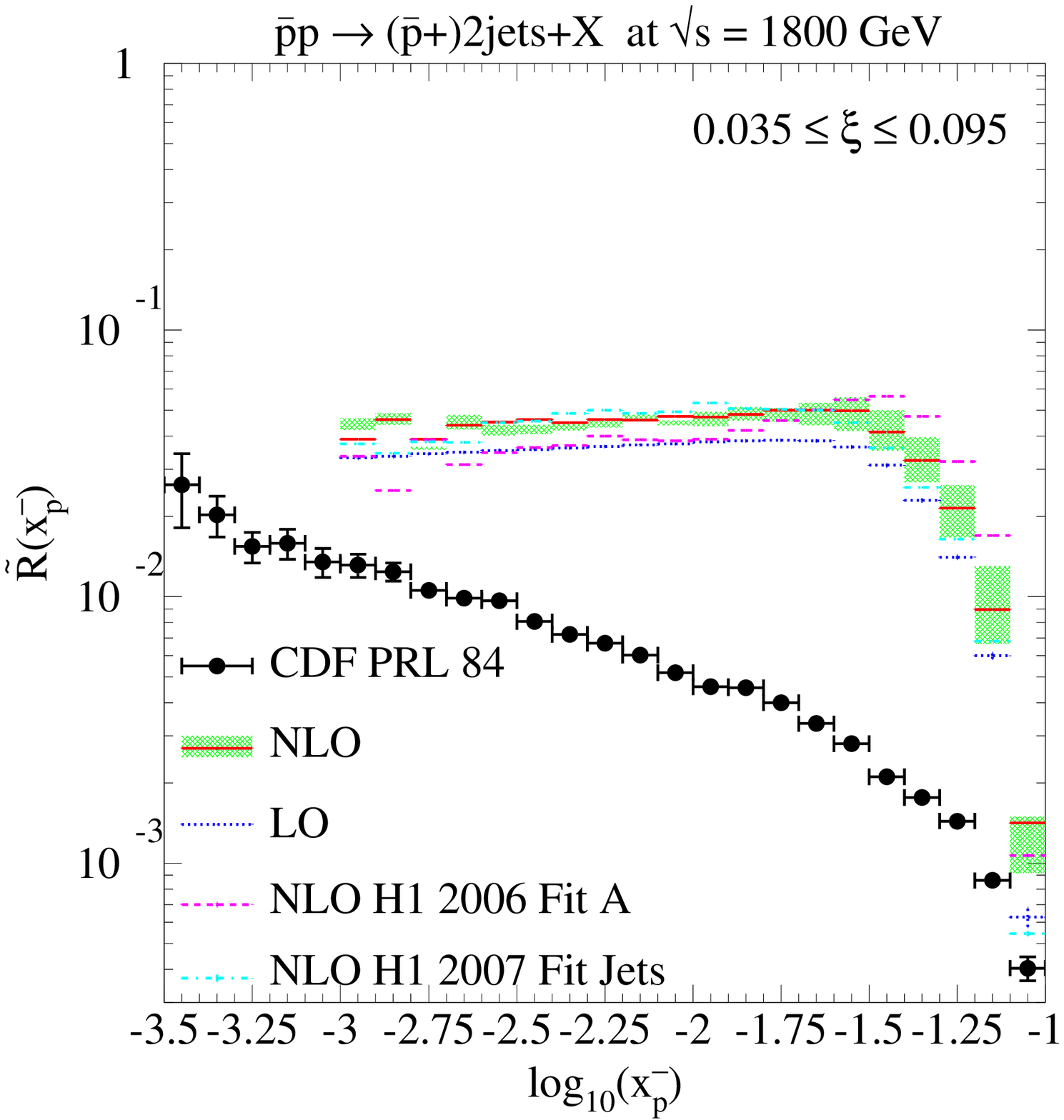}
 \includegraphics[width=0.49\columnwidth]{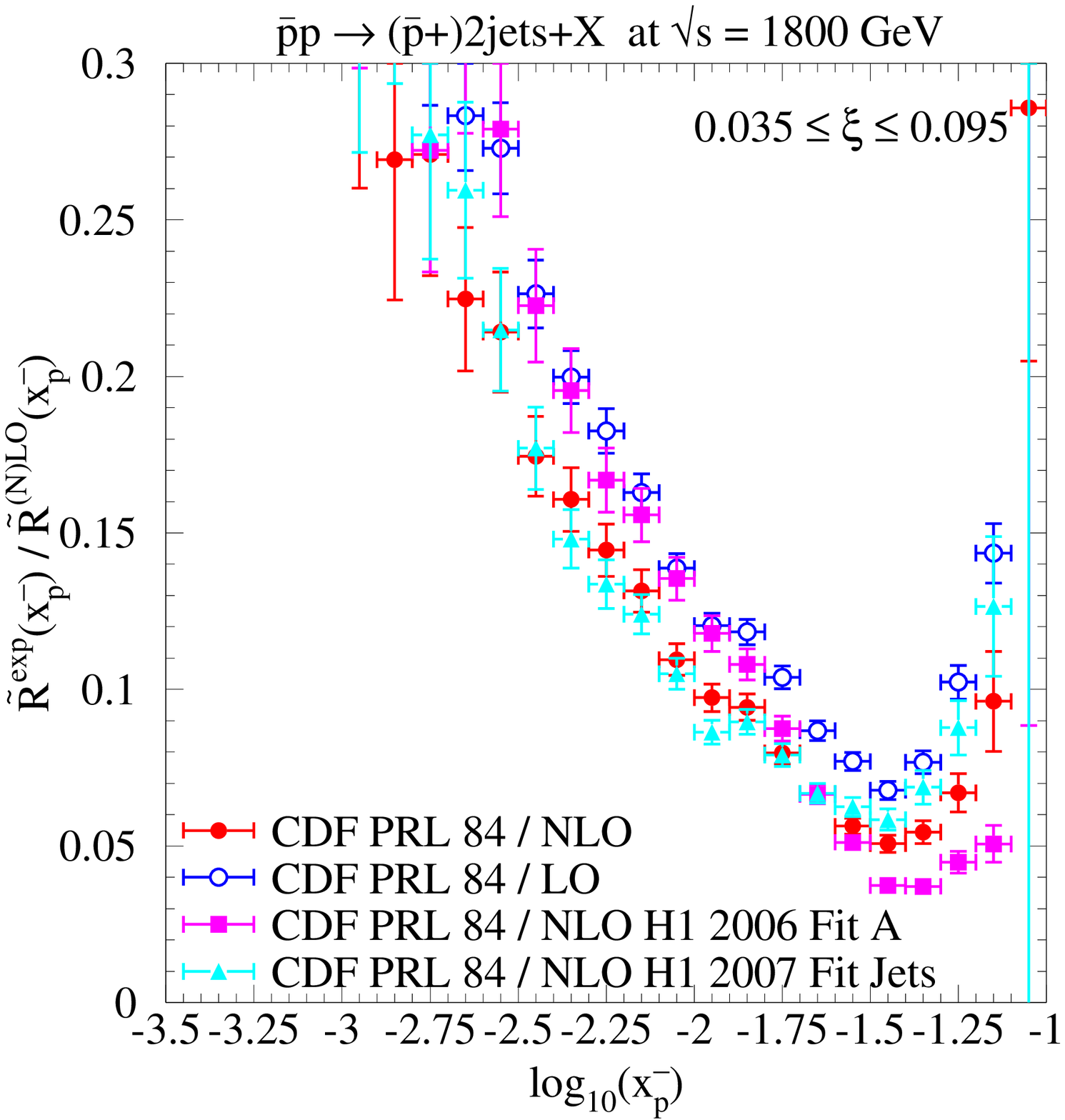}
 \caption{\label{fig:3}Left: Ratio $\R$ of SD to ND dijet cross sections
 as a function of the momentum fraction of the parton in the antiproton,
 computed at NLO (with three different DPDFs) and at LO and compared
 to the Tevatron Run I data from the CDF collaboration. Right: Double
 ratio of experimental over theoretical values of $\R$, equivalent to
 the factorization-breaking suppression factor required for an accurate
 theoretical description of the data (color online).}
\end{figure}
%
The results are plotted in Fig.\ 3 (left) as a function of $\log_{10}
(x_{\bar{p}})$ for three choices of the DPDFs, 'H1 2006 fit B' (full),
'H1 2006 fit A' (dashed) and 'H1 2007 fit jets' (dot-dashed). All three
are calculated in NLO. The NLO scale uncertainty for 'H1 2006 fit B'
(shaded band) cancels out to a large extent in this ratio of cross
sections. The LO prediction for 'H1 2006 fit B' is also
given (dotted). The CDF data, which are plotted in Ref.\ \cite{6} in six
$\xi$ bins of width $\Delta \xi = 0.01$, have been integrated to give
$\R(x_{\bar{p}})$ in the range $0.035 < \xi <0.095$. They were available
in numerical form with statistical errors \cite{22} and are also plotted
in Fig.\ 3 (left). From these presentations it is obvious that the
theoretical ratios $\R(x_{\bar{p}})$ are, depending on $x_{\bar{p}}$, by
up to an order of magnitude larger than the measured $\R(x_{\bar{p}})$ in
agreement with the result in \cite{6}. There are quite some differences
for the different DPDF choices. In general 'fit B' and 'fit jets' lie
closely together, whereas 'fit A' deviates more or less from these two
depending on the $x_{\bar{p}}$ range. For 'fit B' we also show the
scale error and the LO prediction. The hierarchy between the three
DPDFs at large $x_{\bar{p}}$ is easily explained by the fact that the 
corresponding gluon DPDFs are at large $x_{\bar{p}}$ the largest for
'fit A' and the smallest for 'fit jets' \cite{4,20}. The same pattern
between the different DPDFs is seen even more clearly if we plot the
ratio of the experimental $\R(x_{\bar{p}})$ and the theoretical
$\R(x_{\bar{p}})$ as a function of $\log_{10}(x_{\bar{p}})$. The result
for this (double) ratio $\R^{\rm exp}(x_{\bar{p}})/\R^{\rm (N)LO}
(x_{\bar{p}})$
is seen in Fig.\ 3 (right). As can be seen, this ratio varies in a
rather similar way for the three DPDFs in NLO and for 'fit B' in LO
in the range $10^{-3} < x_{\bar{p}} < 10^{-1}$. The variation is
strongest for the 'fit A' DPDF, where this ratio varies by more than
a factor of seven. For the other two DPDFs this variation is somewhat
less, but still appreciable. Actually, we would expect that the ratio
plotted in Fig.\ 3 (right), which gives us the suppression factor,
should vary only moderately with $x_{\bar{p}}$. After presenting the
$\sqrt{s}=630$ GeV and the more recent $\sqrt{s}=1800$ GeV data
below, we shall discuss possible interpretations of this
result. We also observe that the suppression factor for 'fit B' in
NLO and LO are different, in particular for the very small
$x_{\bar{p}}$.

In Ref. \cite{6}, the ratio $\R(x_{\bar{p}})$ was multiplied with an
effective PDF governing the ND cross section to obtain the effective
DPDF $\F_{\rm JJ}^{\rm D}(\beta)$ as a function of $\beta=x_{\bar{p}}/\xi$.
This effective non-diffractive PDF $F_{\rm JJ}^{\rm ND}(x)$ is calculated from
the formula in Eq.\ (2), where the gluon PDF $g(x)$ and the quark PDFs
$q_i(x)$ are taken from the GRV98 LO parton density set \cite{23} and
evaluated at the scale $Q^2=75$ GeV$^2$, corresponding to the average
$E_T^{\rm jet}$ of the SD and ND jet cross sections. Then, the
effective diffractive PDF $\F_{\rm JJ}^{\rm D}(\beta)$ of the antiproton is
obtained from the equation
\beq
 \F_{\rm JJ}^{\rm D}(\beta) = \R(x=\beta\xi) \times
 \F_{\rm JJ}^{\rm ND}(x \to \beta\xi).
\eeq
We use this relation for the experimental and theoretical values of
$\R(x_{\bar{p}})$. However, both are integrated over $\xi$ and are
not given as function of $\xi$. We consider them as only moderately
$\xi$-dependent and evaluate $\R(x=\beta\xi)$ and $\F_{\rm JJ}^{\rm ND}(x\to
\beta \xi)$ at an average value of $\bar{\xi}=0.0631$. This works
quite well over the $\beta$-range of interest, if we compare the
$\F_{\rm JJ}^{\rm D}(\beta)$ values obtained in this way with the
$\F_{\rm JJ}^{\rm D}(\beta)$ in the CDF publication \cite{6}, as is seen
in Fig.\ 4
%
\begin{figure}
 \centering
 \includegraphics[width=0.49\columnwidth]{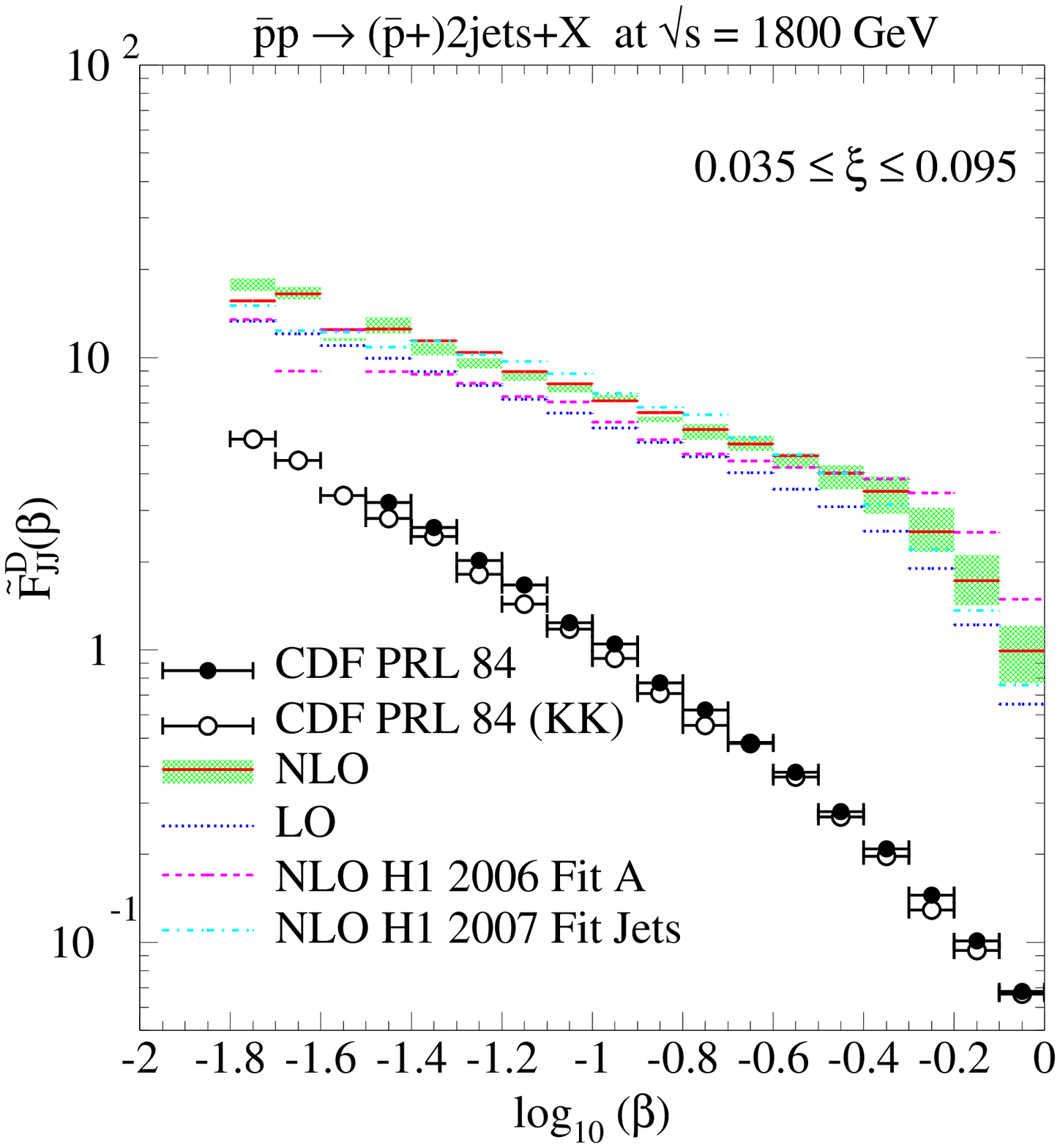}
 \includegraphics[width=0.49\columnwidth]{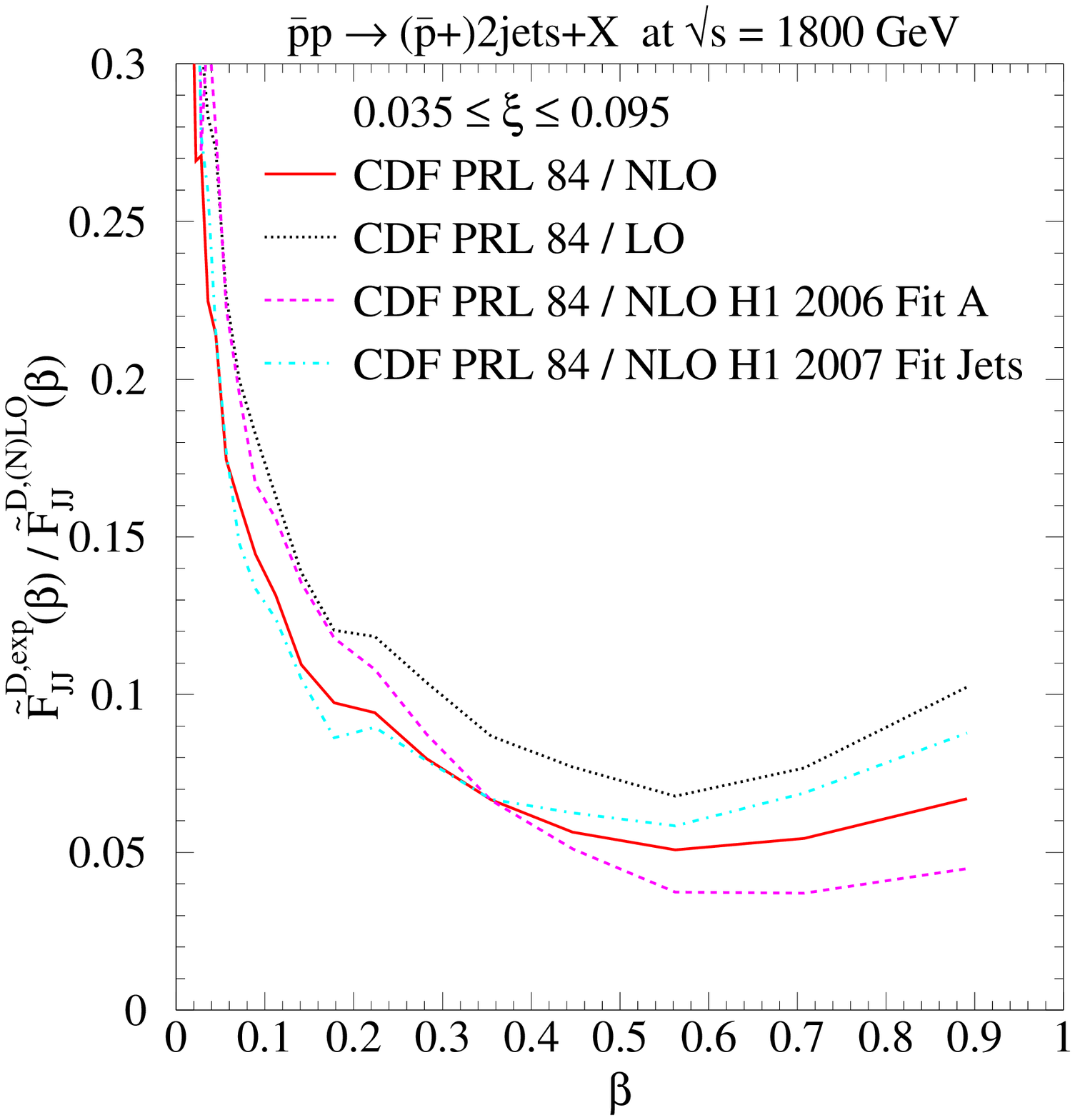}
 \caption{\label{fig:4}Left: Effective diffractive structure function
 $\F_{\rm JJ}^{\rm D}$ of the partons with momentum fraction $\beta$ in the
 Pomeron as measured in dijet production at the Tevatron and compared to
 our (N)LO calculations. Right: Ratio of experimental over theoretical
 values of $\F_{\rm JJ}^{\rm D}$, equivalent to the factorization-breaking
 suppression factor required for an accurate theoretical description of
 the data (color online).}
\end{figure}
%
(left), where the full published points from \cite{6}, denoted CDF
PRL 84, coincide rather well with the open points deduced from the
above equation and the published values of $\R$, denoted CDF
PRL 84 (KK). The ratio $\F_{\rm JJ}^{\rm D,exp}(\beta)/\F_{\rm JJ}^{\rm D,(N)LO}
(\beta)$, plotted in Fig.\ 4 (right) as a function of $\beta$ linearly,
gives the suppression factor as a function of $\beta$ instead of
$x_{\bar{p}}$ as in Fig.\ 3. For example, for the 'fit B' DPDF it varies
between 0.3 at $\beta= 0.05$ and 0.13 at $\beta=0.1$ to 0.07 at
$\beta=0.9$. In the range $0.3<\beta<0.9$ the suppression factor varies
only moderately with $\beta$, but increases strongly for $\beta<0.3$,
independently of the chosen DPDF. Above $\beta=0.3$, 'fit B' and 'fit jets' show
the most constant behavior. Here one should note that the result in Fig.\
4 (right) is independent of the assumptions inherent in Eq.\ (4), since
$F_{\rm JJ}^{\rm ND}(x\to\beta\xi)$ cancels in the ratio. The information in
this figure concerning the suppression factor
is equivalent to Fig.\ 4 of the CDF publication \cite{6}.
The main difference to the CDF plot is the fact that now the suppression
factor is given by comparing to calculated NLO cross sections without
using the approximate formula Eq.\ (4) above, which can be justified
only in LO.

To obtain an idea how large the effect of our NLO dijet evaluation
compared to a simple combination of LO parton densities in the Pomeron
is, we have calculated the ratio $\F_{\rm JJ}^{\rm D,NLO}(\beta)/
\F_{\rm JJ}^{\rm D,LO}(\beta)$ for the three DPDFs. Here the numerator is the
$\F_{\rm JJ}^{\rm D}$ from Eq.\ (4) with $\R$ evaluated in NLO, i.e.\
Fig.\ 4 (left), and the denominator is
$\F_{\rm JJ}^{\rm D,LO}(\beta)$ calculated from the formula
\bea
 \F_{\rm JJ}^{\rm D,LO}(\beta) =
 \int\d\xi\int\d t f_{\p/\bar{p}}(\xi,t)
 \beta [g(\beta)+\frac{4}{9} \sum_{i} q_i(\beta)],
\eea
where the Pomeron flux factor $f_{\p/\bar{p}}(\xi,t)$ and the
gluon and quark PDFs in the Pomeron $g(\beta)$ and $q_i(\beta)$
are taken from the fits 'H1 fit A',
'H1 fit B' and 'H1 fit jets' at the scale $Q^2=75$ GeV$^2$,
respectively.
%
%
At $\beta=0.1$, we obtain ratios of 0.95, 1.05 and 1.1 for these three
fits, respectively, indicating
that our more accurate NLO calculations lead to very similar suppression
factors as the simple approximation in Eq.\ (5) for all three DPDFs.
This ratio is more or
less constant as a function of $\beta$ in the considered range, meaning
that already in the CDF publication \cite{6} one has the strong
variation of the suppression factor with $\beta$ as mentioned above.
It is interesting to note that replacing the approximate Eq.\
(5) with the experimentally used Eq.\ (4) compensates the effect
of the NLO corrections, as the ratio of SD to ND $K$-factors, or
equivalently the ratio of the NLO over the LO value of $\R$, is
approximately $1.35$ for the $1800$ GeV calculation discussed here
and $1.6$ for the $630$ GeV calculation presented in the next
subsection. To compute the effect of this approximation alone, i.e.\
the ratio of Eq.\ (4) at LO over Eq.\ (5), one must divide the 
values of 0.95, 1.05 and 1.1 by the ratio of $K$-factors, i.e.\ 1.35.


\subsection{Comparison with 630 GeV and new 1800 GeV data}

In a second publication, the CDF collaboration presented data for
diffractive and non-diffractive jet production at $\sqrt{s}=630$ GeV and
compared them with a new measurement at $\sqrt{s}=1800$ GeV \cite{16}.
From both measurements they deduced diffractive structure functions
using the formula Eq.\ (4) with the expectation that $\F_{\rm JJ}^{\rm D}
(\beta)$ is larger at $\sqrt{s}=630$ GeV than at $\sqrt{s}=1800$ GeV.
The experimental cuts are similar to the cuts in the first analysis
\cite{6} with the exception that now $(-t) \leq 0.2$ GeV$^2$ and in
addition to the $E_T^{\rm jet1,2} > 7.0$ GeV cut they require
$\overline{E_T} > 10$ GeV. This second cut on $\overline{E_T}$ is very
important for the comparison with the NLO predictions, since with this
additional constraint the infrared sensitivity is not present anymore.

\begin{figure}
 \centering
 \includegraphics[width=0.49\columnwidth]{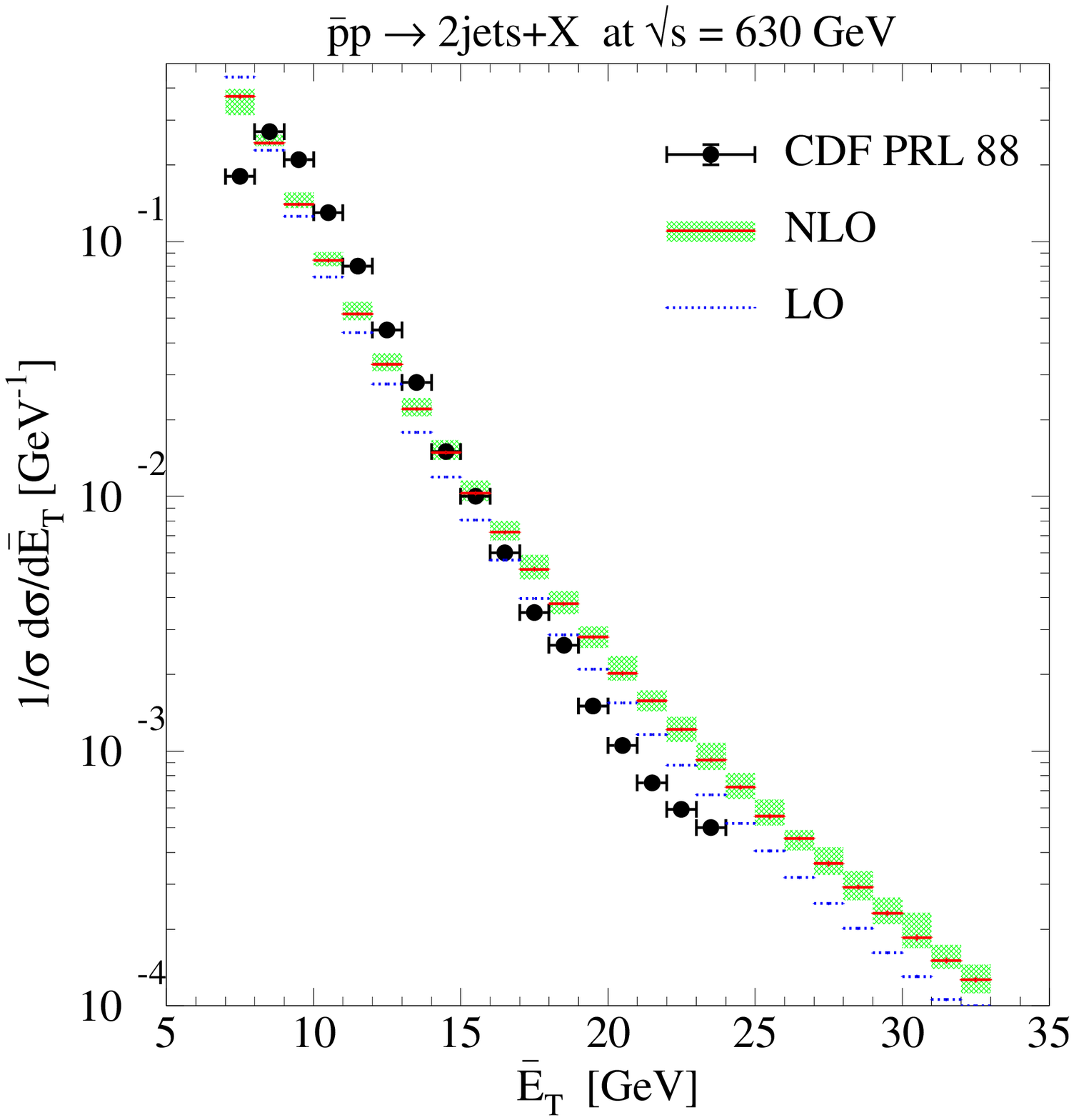}
 \includegraphics[width=0.49\columnwidth]{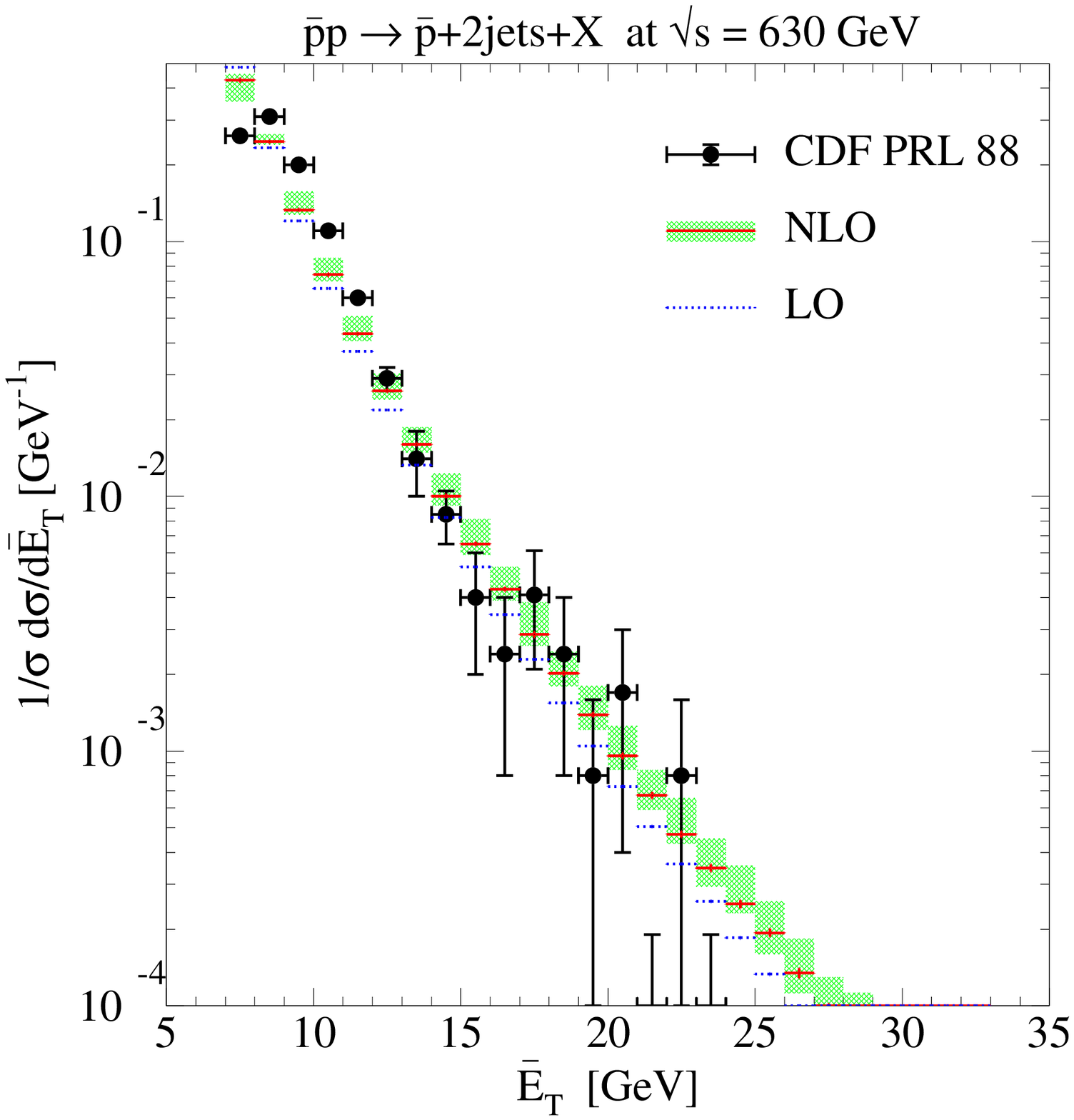}
 \caption{\label{fig:6}Same as Fig.\ 1, but for a reduced center-of-mass
 energy of 630 GeV at the Tevatron (color online).}
\end{figure}

With these cuts and the integration over $\xi$ in the range $0.035 <
\xi < 0.095$, we have calculated the normalized cross sections
$(1/\sigma)\d\sigma /\d\overline{E_T}$ and  $(1/\sigma)\d\sigma/\d
\overline{\eta}$ as in the previous subsection, but now for
$\sqrt{s}=630$ GeV. For ND (left) and SD (right) jet production, the
results are shown in Fig.\ \ref{fig:6} and compared to the data from Ref.\
%
%
\cite{16}. Here, the ND data sample was larger by about two orders of
magnitude compared to the SD data sample, so that the statistical errors,
which were not given in Ref.\ \cite{16}, should be smaller by about a
factor of ten \cite{20a}. Again, no information about systematic errors
was available.
We find reasonably good agreement in the medium-$\overline{E_T}$
range. In these figures, we have also plotted the LO predictions
(dotted). For the DPDF, we have chosen as before the 'H1 2006 fit B'
set. Due to the large experimental errors for $\overline{E_T} > 15$ GeV
for the SD case, we also find good agreement in the large-$\overline{E_T}$
range.

\begin{figure}
 \centering
 \includegraphics[width=0.49\columnwidth]{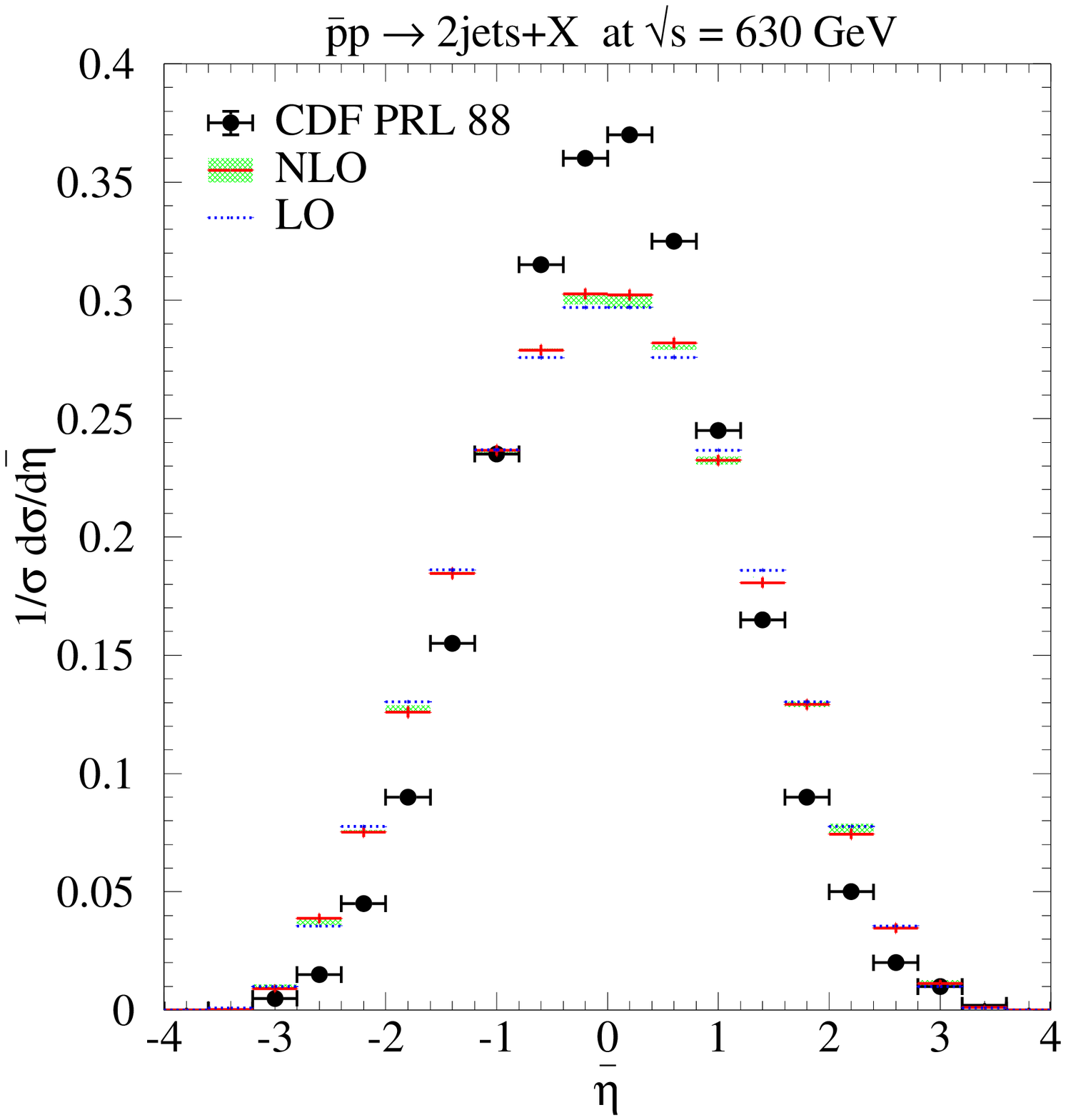}
 \includegraphics[width=0.49\columnwidth]{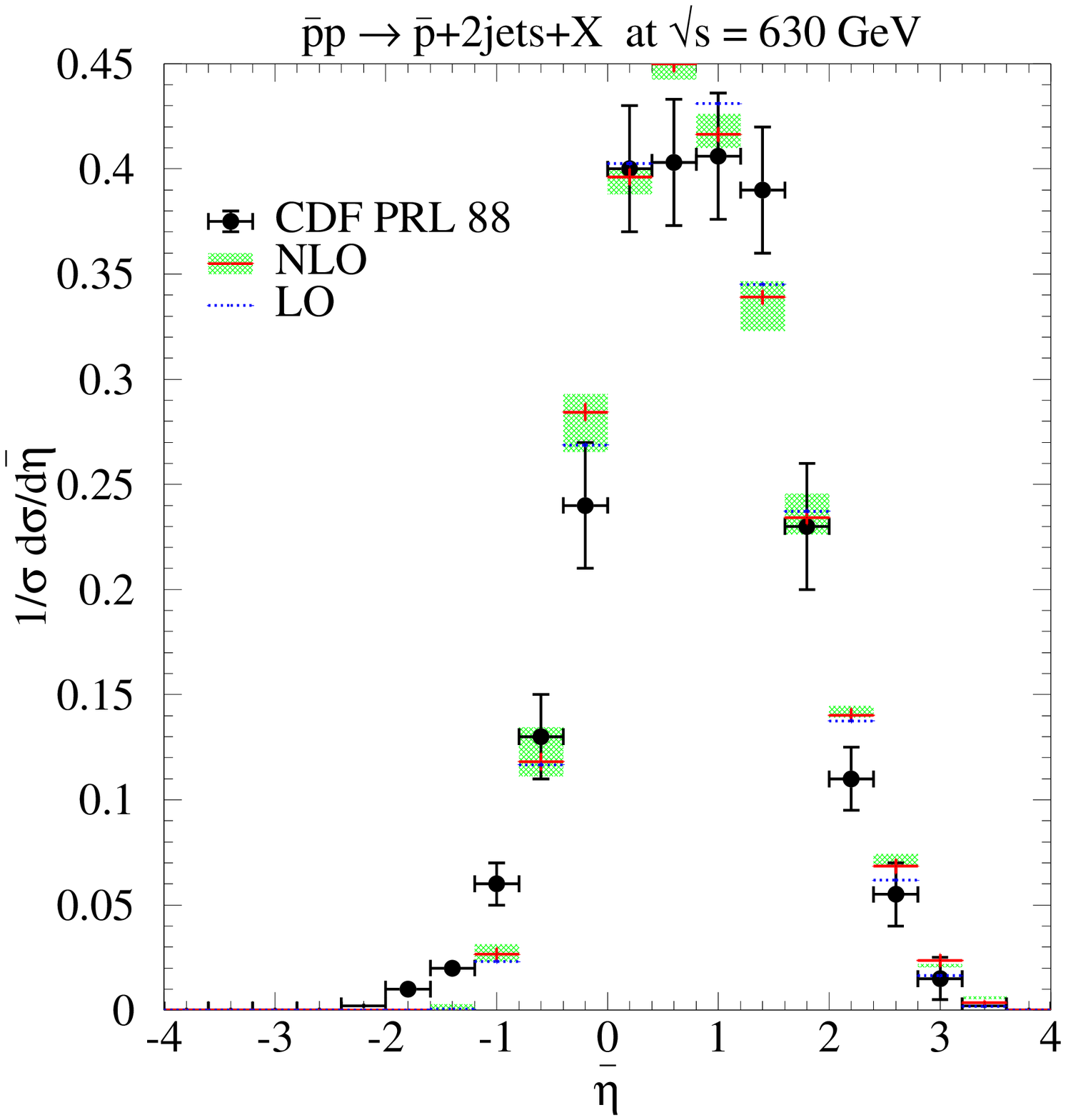}
 \caption{\label{fig:7}Same as Fig.\ 2, but for a reduced center-of-mass
 energy of 630 GeV at the Tevatron (color online).}
\end{figure}

The equivalent result and comparison with the data for the
$\overline{\eta}$-distribution is shown in Fig.\ \ref{fig:7}, again for ND (left)
%
%
and SD (right) jet production for NLO (full) and LO (dotted)
predictions. The agreement between the theoretical results and the CDF
data is similar as in the previous subsection, where we compared to the
$\sqrt{s}=1800$ GeV data. This justifies to go on with the calculation
of the ratio $R(x_{\bar{p}})$ of SD to ND cross sections.

\begin{figure}
 \centering
 \includegraphics[width=0.49\columnwidth]{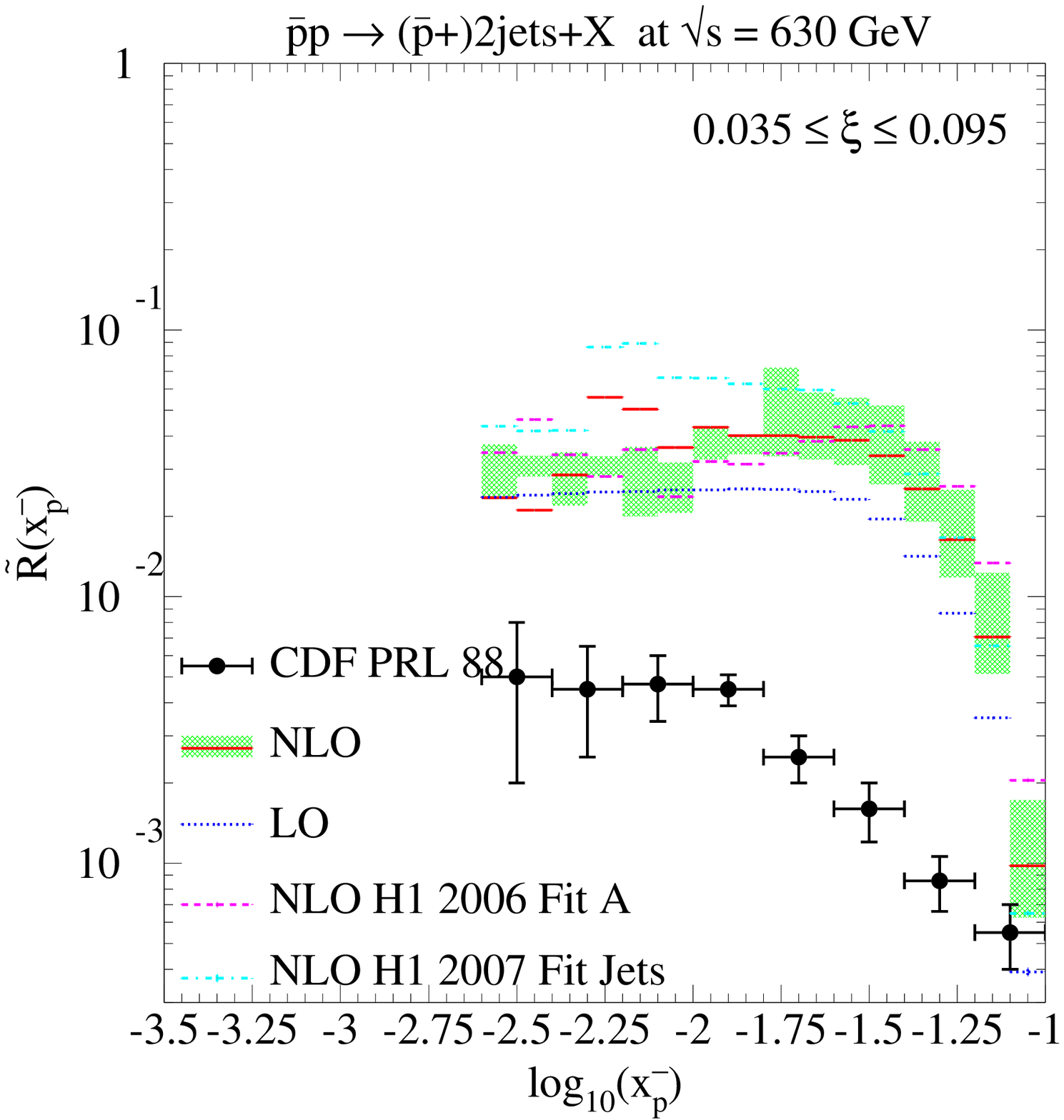}
 \includegraphics[width=0.49\columnwidth]{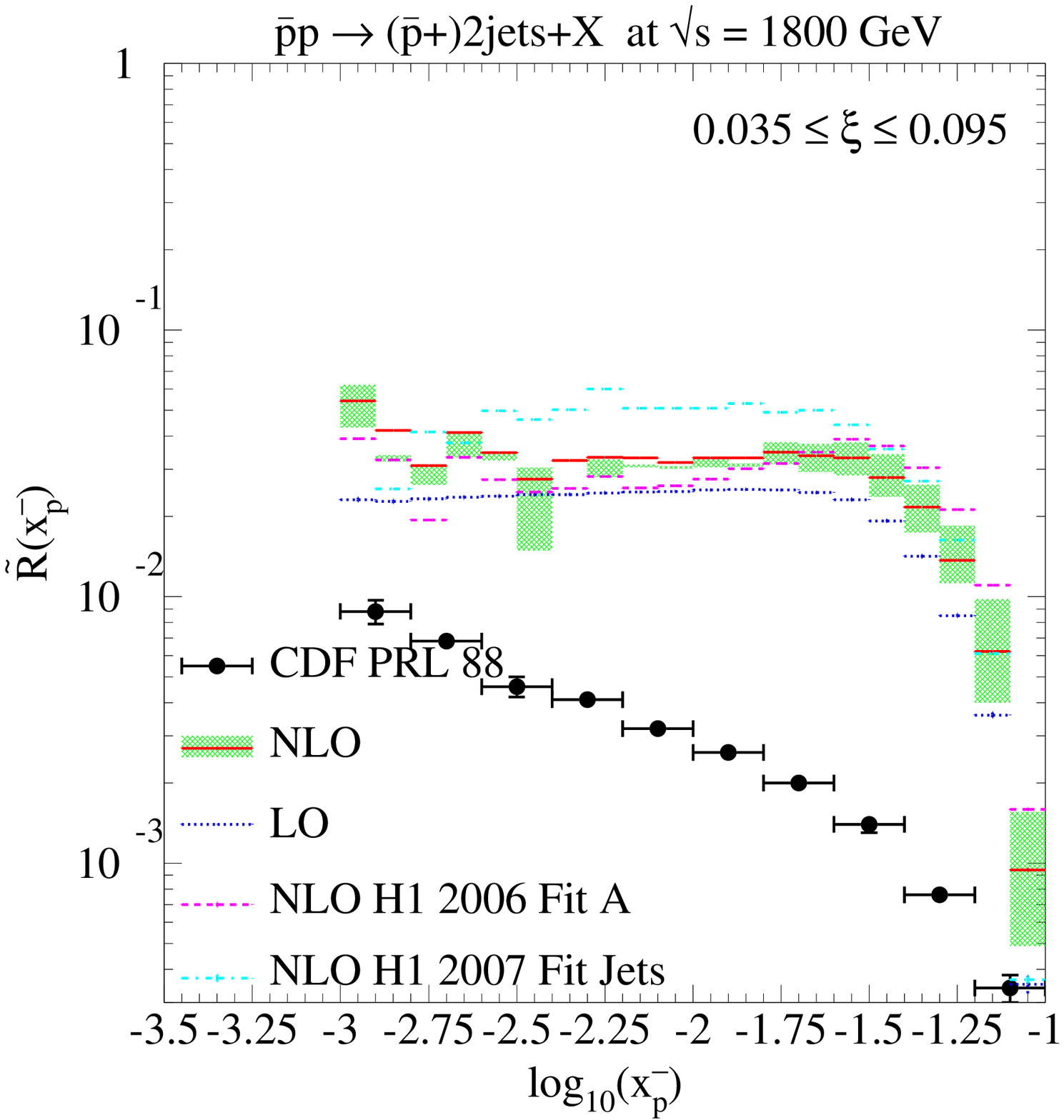}
 \caption{\label{fig:8}Ratios $\R$ of SD to ND dijet cross sections
 as a function of the momentum fraction of the parton in the antiproton,
 computed at NLO (with three different DPDFs) and at LO and compared
 to the Tevatron data at $\sqrt{s}=630$ (left) and 1800 GeV (right) from
 the CDF collaboration (color online).}
\end{figure}

The momentum fraction $x_{\bar{p}}$ is calculated as before from Eq.\
(3), and then the cross sections $\d\sigma/\d x_{\bar{p}}$ can be
calculated with the same restrictions on the number of included jets as
before. The only difference is the different cut on $(-t)$. The results
for $\R(x_{\bar{p}})$ at $\sqrt{s}=630$ GeV are presented in Fig.\ \ref{fig:8}
%
%
(left) for the three choices of
DPDFs, 'fit B' (NLO and LO), 'fit A' (NLO) and 'fit jets' (NLO). In this
figure, also the experimental data from Ref.\ \cite{16} are included.
The range of $x_{\bar{p}}$ is now much smaller than for the $\sqrt{s}=
1800$ GeV case. It ranges from $x_{\bar{p}} = 0.025$ to $x_{\bar{p}}=
0.1$. From this plot, the suppression of the SD cross section is clearly
visible. The suppression factor is of the same order of magnitude as in
the previous subsection. The same plot for the new $\sqrt{s}=1800$ GeV
data \cite{16} together with the predictions is given in Fig.\ \ref{fig:8}
(right).

From the two plots in Fig.\ \ref{fig:8} we have calculated the corresponding
suppression factors $\R^{\rm exp}(x_{\bar{p}})/\R^{\rm (N)LO}
(x_{\bar{p}})$, exhibited in Fig.\ \ref{fig:9} (left: $\sqrt{s}=630$ GeV; right:
%
\begin{figure}
 \centering
 \includegraphics[width=0.49\columnwidth]{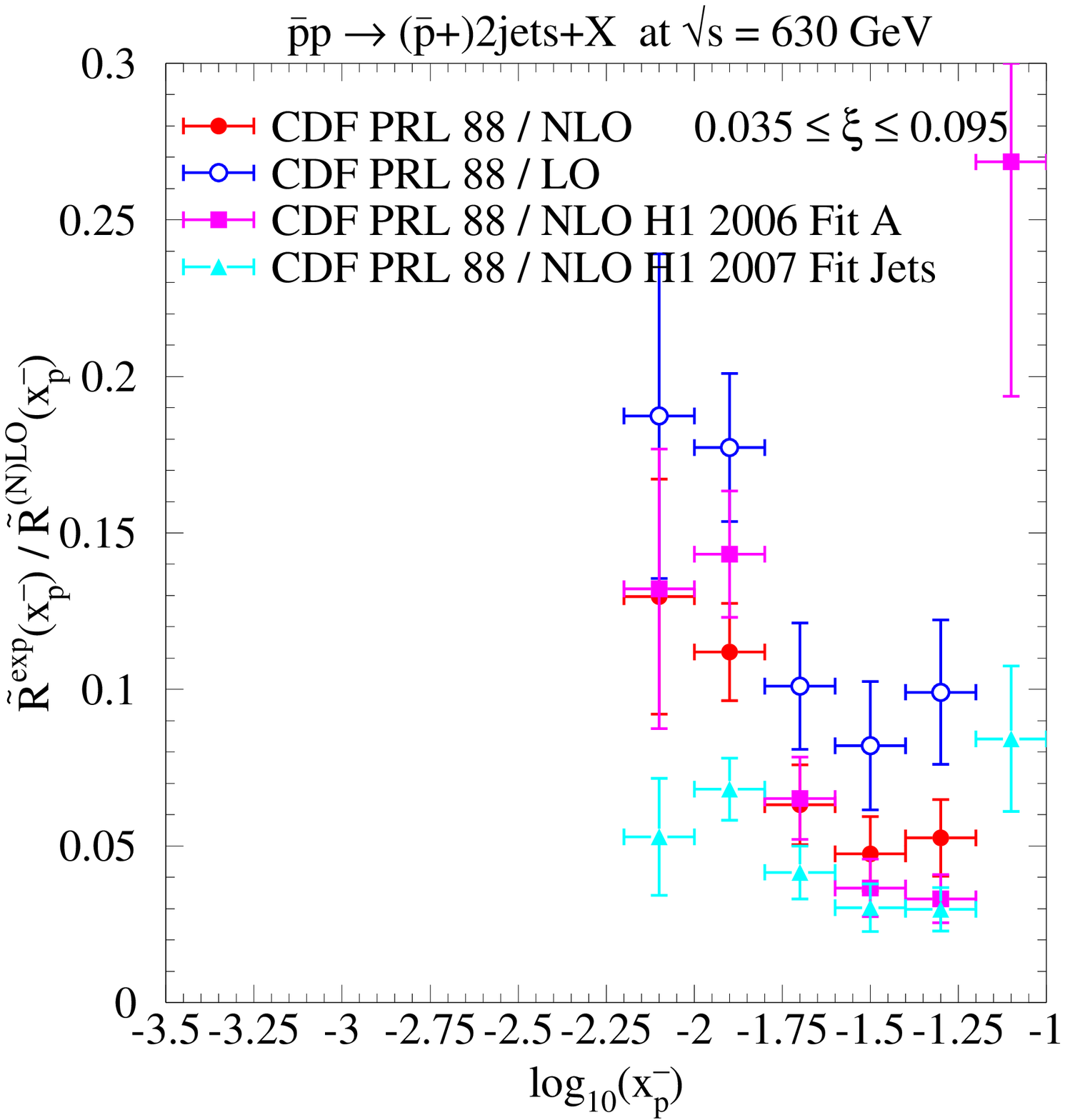}
 \includegraphics[width=0.49\columnwidth]{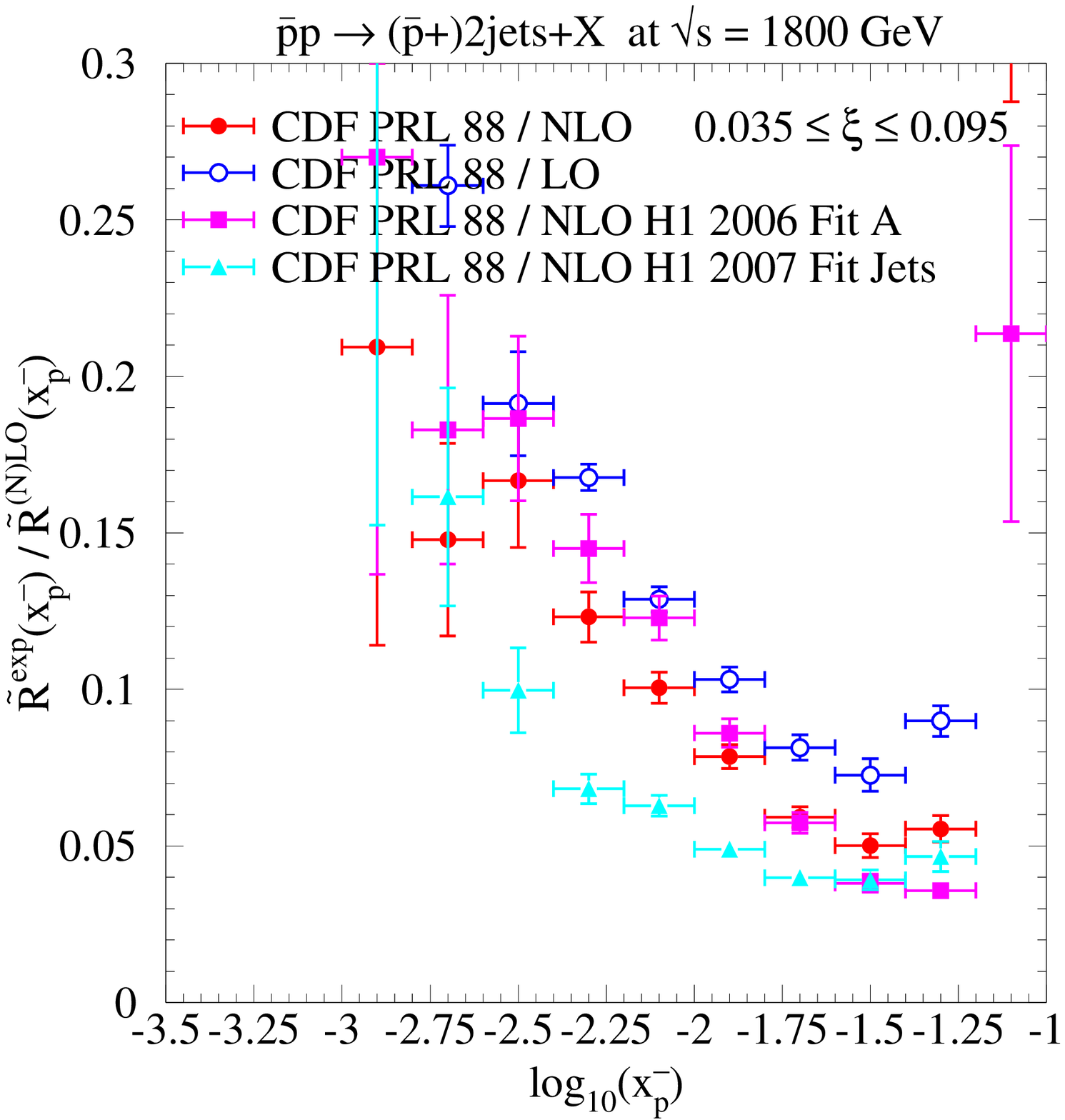}
 \caption{\label{fig:9}Double ratios of experimental over theoretical
 values of $\R$, equivalent to the factorization-breaking suppression
 factor required for an accurate theoretical description of the data
 from the Tevatron at $\sqrt{s}=630$ (left) and 1800 GeV (right)
 (color online).}
\end{figure}
%
$\sqrt{s}=1800$ GeV). In both figures we observe that the LO and NLO
results for the suppression factors differ significantly (LO only given
for 'fit B'), but also the three different DPDFs give different
suppression factors, although with smaller variation compared to the LO
and NLO result. Due to the variation of this factor with $x_{\bar{p}}$
it is difficult to compare the suppression of the $\sqrt{s}=630$ GeV
result (left) with the $\sqrt{s}=1800$ GeV result (right) in Fig.\ \ref{fig:9}.
On average, it seems that for larger $x_{\bar{p}}$ the two suppression
factors are more or less equal and we cannot say that the suppression factor
for $\sqrt{s}=630$ GeV is larger than for $\sqrt{s}=1800$ GeV, as we
would expect it. In the region $x_{\bar{p}} \geq 0.02$, the suppression
factors for both $\sqrt{s}$ are fairly constant ($\simeq 0.05$), in
particular for the DPDF 'fit jets'. This is not the case for the
analysis in the previous subsection, where, as we see in Fig.\ 4, the
suppression factor varies already much more in this particular
$x_{\bar{p}}$ range.

\begin{figure}
 \centering
 \includegraphics[width=0.49\columnwidth]{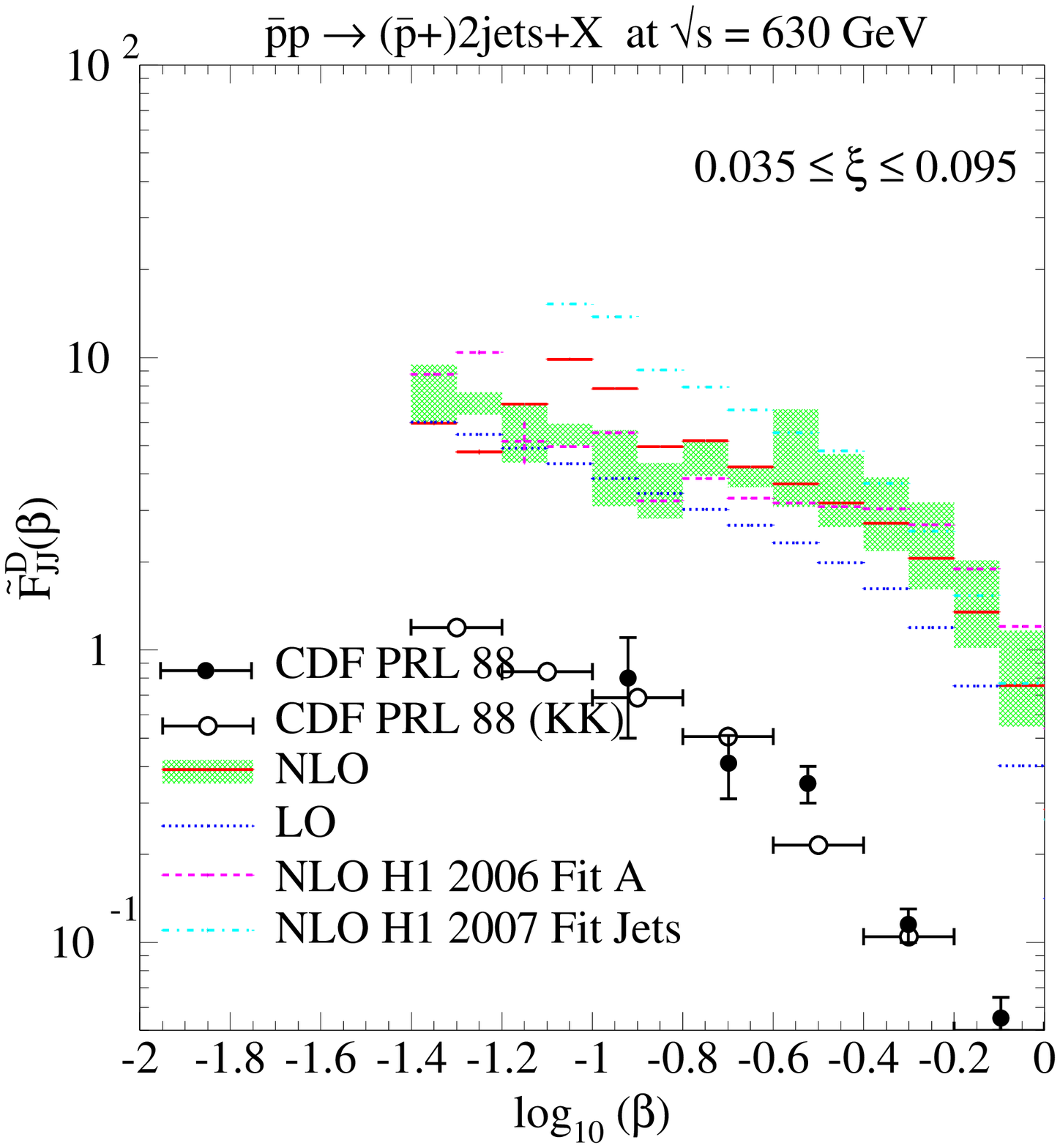}
 \includegraphics[width=0.49\columnwidth]{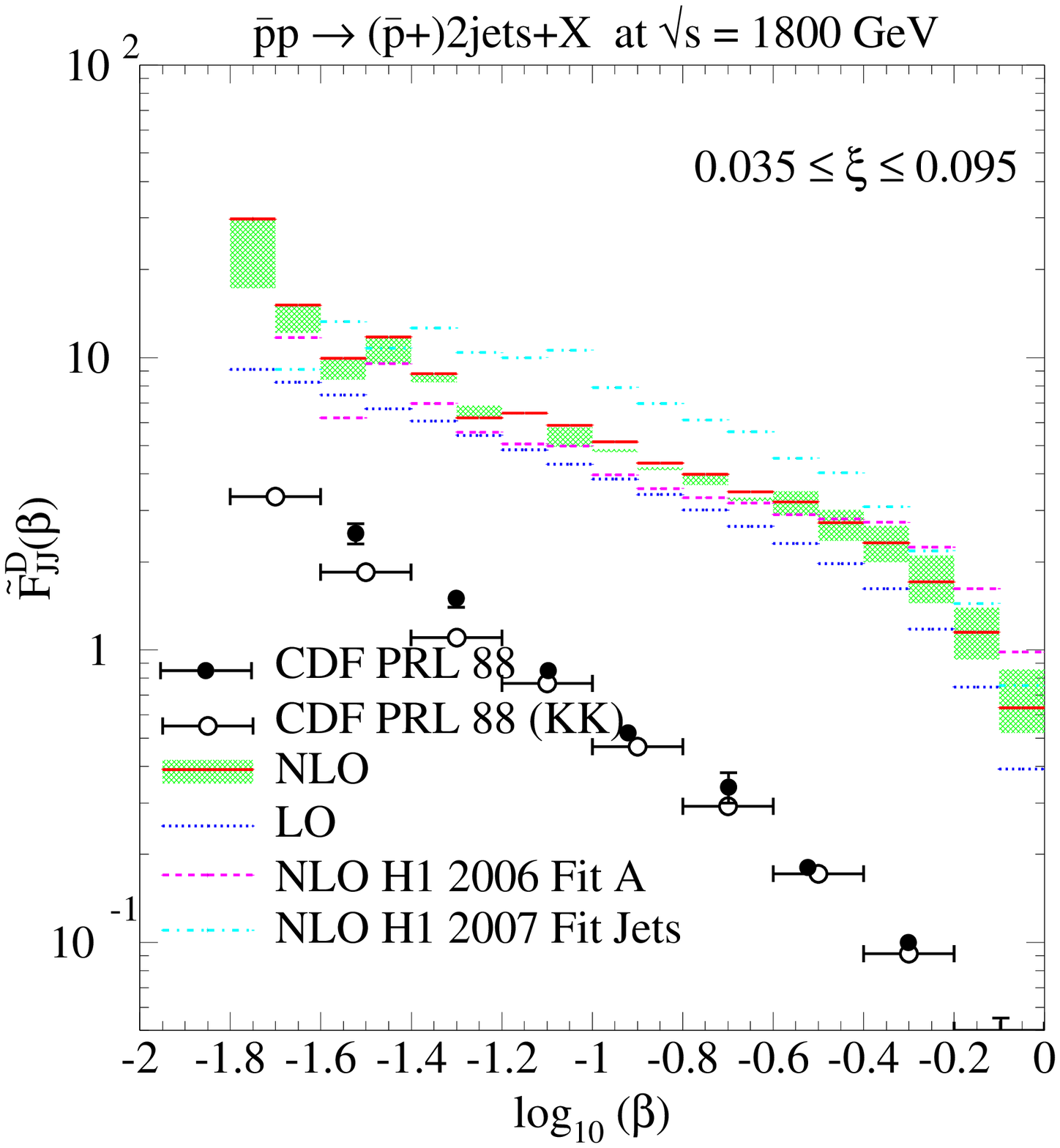}
 \caption{\label{fig:10}Effective diffractive structure function
 $\F_{\rm JJ}^{\rm D}$ of the partons with momentum fraction $\beta$ in the
 Pomeron as measured in dijet production at the Tevatron
 with $\sqrt{s}=630$ (left) and 1800 GeV (right) and compared to
 our (N)LO calculations (color online).}
\end{figure}

From the results in Fig.\ \ref{fig:8}, we have calculated $\F_{\rm JJ}^{\rm D}
(\beta)$ by changing variables from $x_{\bar{p}}$ to $\beta$ with
$\beta=x_{\bar{p}}/\bar{\xi}$ and $\bar{\xi} = 0.0631$ and multiplying
with the effective PDF for ND jet production as in Eq.\ (4). The
results, together with the corresponding experimental data from Ref.\
\cite{16} and those calculated with the chosen $\bar{\xi}$, which agree
inside errors except for two points at small $\beta$, are shown in
Fig.\ \ref{fig:10} (left: $\sqrt{s}=630$ GeV, right: $\sqrt{s}=1800$ GeV).
%
%

\begin{figure}
 \centering
 \includegraphics[width=0.49\columnwidth]{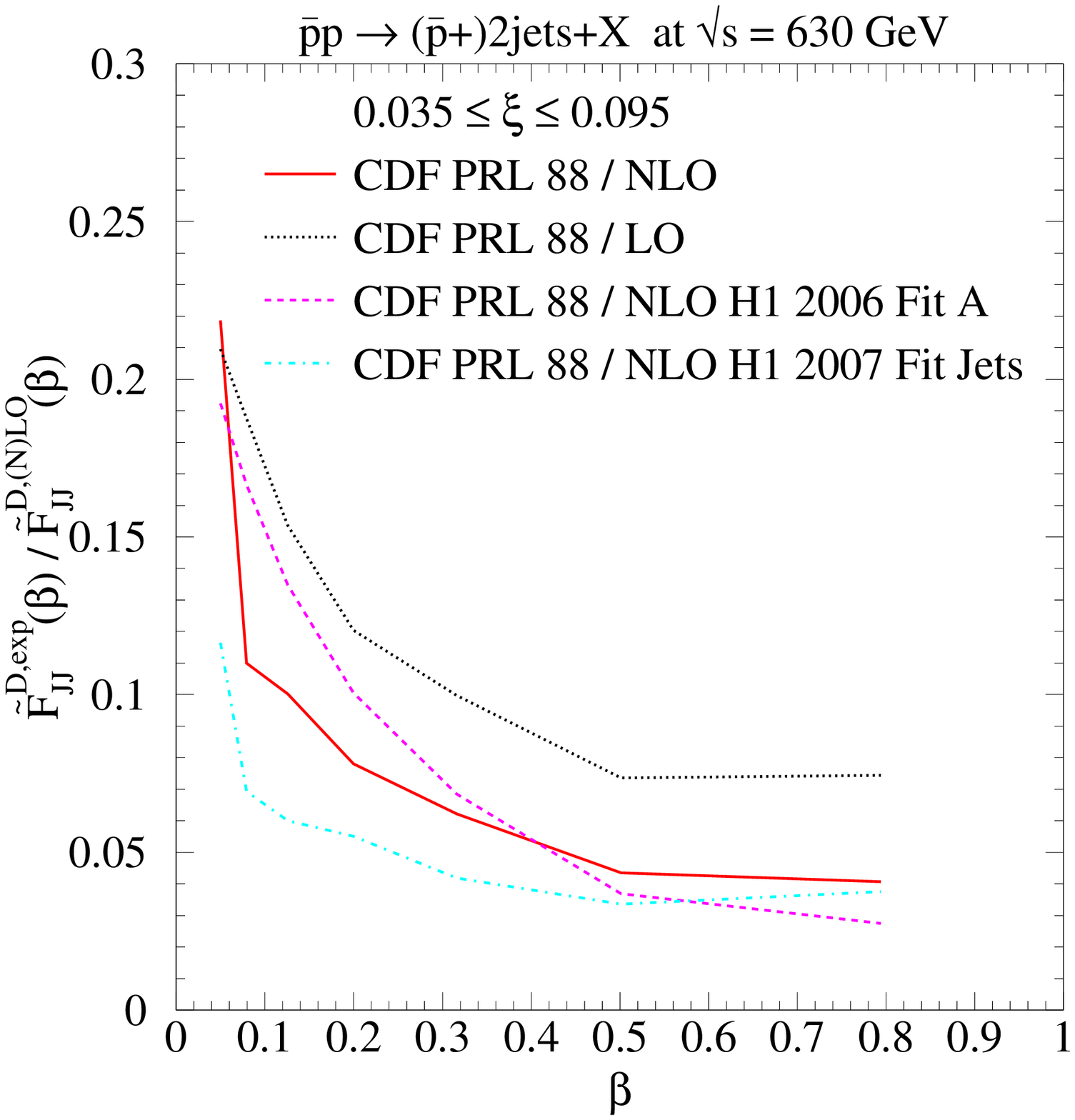}
 \includegraphics[width=0.49\columnwidth]{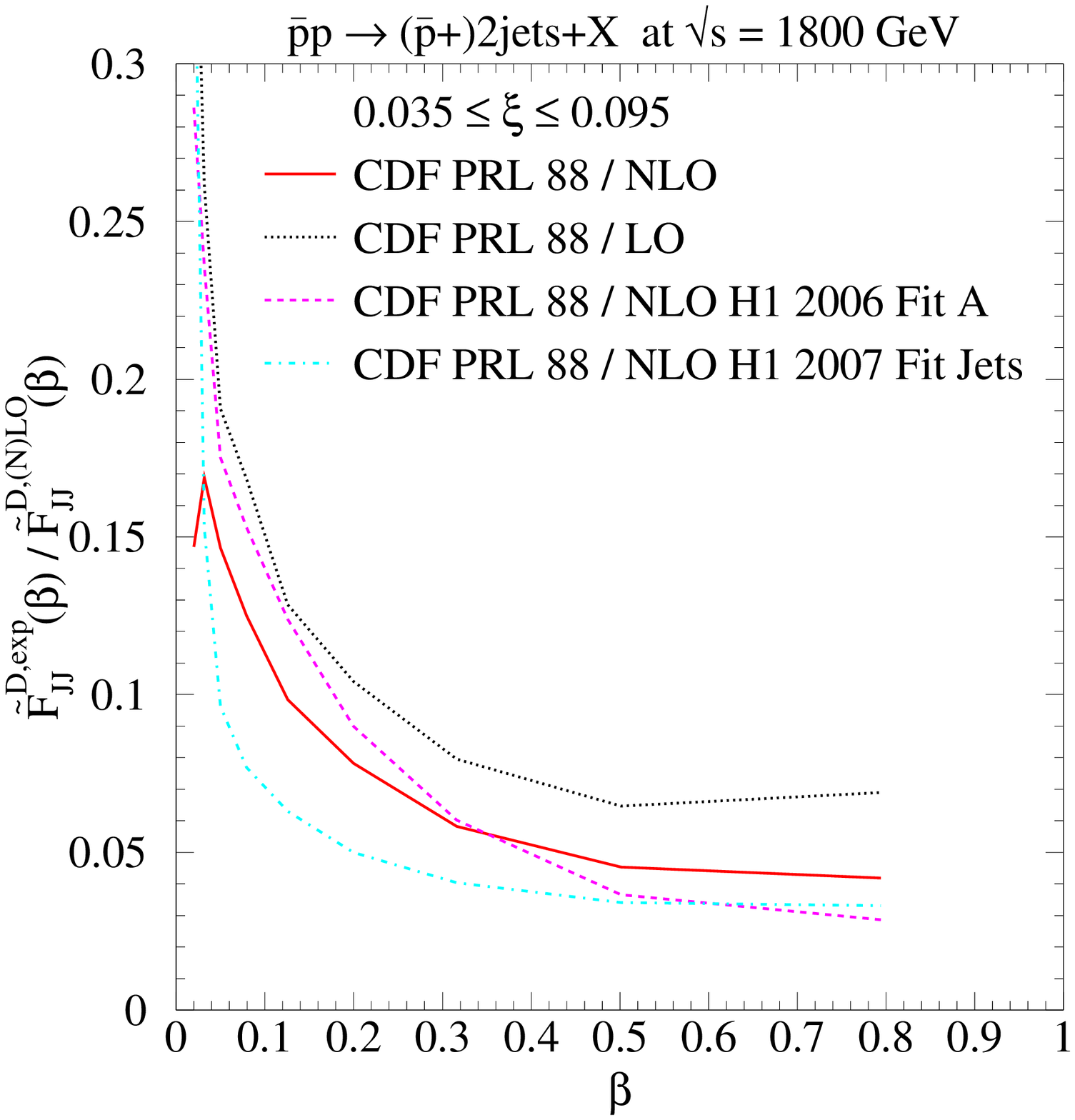}
 \caption{\label{fig:11}Ratios of experimental over theoretical
 values of $\F_{\rm JJ}^{\rm D}$ for $\sqrt{s}=630$ (left) and 1800 GeV
 (right), equivalent to the factorization-breaking
 suppression factors required for an accurate theoretical description of
 the data (color online).}
\end{figure}

From these results we have again calculated, as in the previous
subsection, the suppression factor as a function of $\beta$ in the
range $0<\beta<0.8$. The plots for the ratios $\F_{\rm JJ}^{\rm exp}/
\F_{\rm JJ}^{\rm (N)LO}$ are seen in Fig.\ \ref{fig:11} for the lower (left) and
%
%
the higher center-of-mass energy data (right), again for the three DPDF
fits in NLO and 'fit B' also in LO. First we observe that the ratios
in Fig.\ \ref{fig:11} differ very little, except perhaps at very small $\beta$.
This means that from these data there is no essential difference seen
in the suppression at $\sqrt{s}=630$ GeV and $\sqrt{s}=1800$ GeV.
Second, we notice that with the 'fit jets' we have the most constant
behavior of the suppression for $\beta > 0.2$. Furthermore, comparing
Fig.\ \ref{fig:11} (right) with Fig.\ 4 (right) we see some differences. While the
general pattern is the same, the suppression factor for 'fit jets' in
particular is less constant and larger in Fig.\ 4 (right) than in Fig.\
\ref{fig:11} (right), which is obviously correlated with the more restrictive
cuts on $\overline{E_T}$ and $t$ in the latter.

For completeness we also compared our NLO dijet calculation to the
approximate LO formula in Eq.\ (5).
%
%
For $\sqrt{s}=1800$ GeV and $\beta=0.1$, we obtain the same values of
0.95, 1.05, and 1.1 for 'H1 2006 fit A', 'H1 2006 fit B', and 'H1 2007
fit jets' as for the older CDF analysis. They depend again weakly on
$\beta$. For $\sqrt{s}=630$ GeV and $\beta=0.1$ we obtain larger
values of 1.15, 1.35, and 1.45, which is in line with the larger ratio
of $K$-factors (1.6 instead of 1.35) for SD and ND events at this lower
center-of-mass energy.

\begin{figure}
 \centering
 \includegraphics[width=0.98\columnwidth]{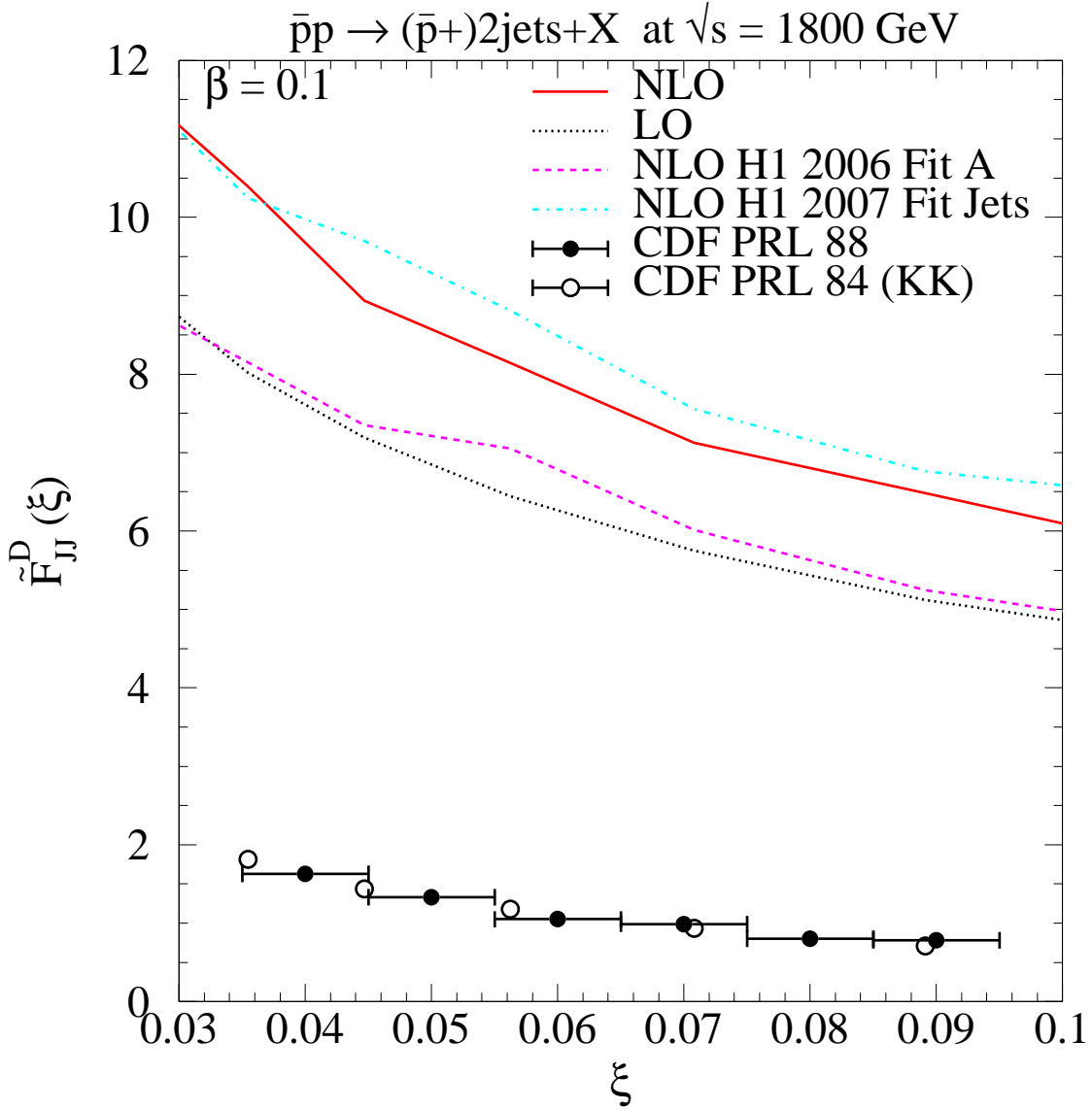}
 \caption{\label{fig:13}Effective diffractive structure function in Eq.\
 (4) from (N)LO dijet cross sections for fixed $\beta$ as a function
 of the momentum fraction of the Pomeron in the antiproton $\xi$,
 compared to the Tevatron data at $\sqrt{s}=1800$ GeV from the CDF
 collaboration \cite{16} (color online).}
\end{figure}

As stated above, the calculation of the effective diffractive structure
function $\F^D_{\rm JJ}(\beta)$ from the ratio $\R(x_{\bar{p}})$ was based
on the assumption that the latter was only weakly $\xi$-dependent, so
that Eq.\ (4) could be evaluated at an average value of $\bar{\xi}=0.0631$.
This weak $\xi$-dependence is indeed observed in the newer CDF data,
published in the lower part of Fig.\ 4 of Ref.\ \cite{16} and
reproduced in our Fig.\ \ref{fig:13} (full circles). These data agree well with
%
%
the $\xi$-dependent values of $\R(x_{\bar{p}})$ published in Fig.\ 3 of
Ref.\ \cite{6} when transformed into $\F^D_{\rm JJ}(\xi)$ using Eq.\ (4)
and $\xi=x_{\bar{p}}/\beta$ with $\beta=0.1$ (open circles). The same
weak $\xi$-dependence is also observed in our theoretical calculations
when using the same procedure, except with different normalization,
reflecting the $\xi$-dependence of the H1 fits to the Pomeron
flux factors $f_{\p/\bar{p}}(\xi,t)\propto \xi^{-m}$ with 
$m\simeq 1.1$ (0.9 in the CDF fit to their data). At
the considered
value of $\beta=0.1$, the NLO suppression factors for 'fit A,B' and
'fit jets' are 0.15, 0.12 and 0.11, respectively, and are almost independent
of $\xi$. At LO, the suppression factor for 'fit B' is larger, i.e.\
0.15, which corresponds to the fact that the ratio of SD over ND
$K$-factors is 1.35. Note that Fig.\ \ref{fig:13} is based on the higher statistics
CDF data without the stronger cuts on $\overline{E_T}$ and $t$ and should
therefore be not compared to Fig.\ \ref{fig:11} (right), but to Fig.\ 4 (right),
where consistency of these numbers with the values shown at $\beta=0.1$
can be found.

The (small) difference of the theoretical (1.1) and experimental (0.9)
values of $m$ can be explained by a subleading Reggeon contribution, which
has not been included in our predictions. To study its importance, we
have computed the ratio of the Reggeon over the Pomeron contribution to the
LO single-diffractive cross section at $\sqrt{s}=1800$ GeV
as a function of $\overline{E_T}$,
$\bar{\eta}$, and $x_{\bar{p}}$. The Reggeon flux factor was obtained from
'H1 2006 fit B' and convolved, as it was done in this fit, with the parton
densities in the pion of Owens \cite{24}. Very similar results were
obtained for the 'H1 2007 fit jets' Reggeon flux. On average, the Reggeon
adds a 5\% contribution to the single-diffractive cross section, which is
almost independent of $\overline{E_T}$, but is smaller in the proton
direction at large $\bar{\eta}$ and small $x_{\bar{p}}=\xi\beta$ and $\xi$
(2.5\%) than at large $x_{\bar{p}}$ and $\xi$ (8\%). This corresponds to
the graphs shown in Figs.\ 5 ($\xi=0.01$) and 6 ($\xi=0.03$) of the 
H1 publication \cite{4}, e.g.\ at $Q^2=90$ GeV$^2$. While the Reggeon
contribution thus increases the diffractive cross section and reduces
the suppression factor at large $x_{\bar{p}}$ in Figs.\ 3,
8, and 9,
making the latter more constant, the same is less true at small values of 
$x_{\bar{p}}$.


\subsection{Interpretation of the observed suppression factor}

Our main results are the plots for the suppression factors as a function of
$\log_{10}(x_{\bar{p}})$ in Fig.\ 3 (right) deduced from the data of Ref.\
\cite{6} and in Fig.\ \ref{fig:9} from the data of \cite{16} at $\sqrt{s}=630$
GeV (left) and $\sqrt{s}=1800$ GeV (right). The qualitative behavior of the
suppression factor in these three figures is very similar. We observe an
appreciable dependence of the suppression factor on the chosen DPDFs and a
dependence on $x_{\bar{p}}$ with a minimum at $x_{\bar{p}} \simeq 0.032$
($\log_{10}(x_{\bar{p}})\simeq -1.5)$ and a rise towards smaller $x_{\bar{p}}$
by up to a factor of five. The equivalent result as a function of $\beta$ is
shown in Fig.\ 4 (right) for the data of Ref.\ \cite{6} and in Fig.\ \ref{fig:11}
(left) for the $\sqrt{s}=630$ GeV data of Ref.\ \cite{16} and in Fig.\
\ref{fig:11} (right) for the $\sqrt{s}=1800$ GeV data of Ref.\ \cite{16}.
Depending on the chosen DPDFs, the suppression factor as a function of $\beta$ is
minimal with the value $\simeq 0.05$ at $\beta = 0.5$ and rises with decreasing
$\beta$ to a value $\simeq 0.12$ in Fig.\ 4 (right) and Fig.\ \ref{fig:11}
(right) and to a value $\simeq 0.1$ in Fig.\ \ref{fig:11} (left) at $\beta=0.1$
(considering 'fit B' as an example). Of course, this rise of the suppression
factor towards small $\beta$ is directly related to its rise as a function of
$x_{\bar{p}}$ towards small $x_{\bar{p}}$.

A comparison of the H1 data \cite{4}, which are used to obtain the DPDFs applied
in our calculation, with a similar measurement, in which the leading proton is
directly detected \cite{H1LP}, yields a ratio of cross sections for $M_Y < 1.6$
GeV and $M_Y = m_p$ of $1.23\pm 0.03(stat.)\pm 0.16(syst.)$ \cite{4}. Since the
CDF measurements are performed by triggering on the leading antiproton, these
measurements must be multiplied by this ratio to normalize them to the $M_Y<1.6$
GeV constraint for the H1 DPDFs. Therefore, all suppression factors obtained so
far must be multiplied by this ratio.

Any model calculation of the suppression factor, which
is also sometimes called the rapidity gap survival factor, must try to explain
two points, first the amount of suppression, which is $\simeq 0.1$ at $\beta=0.1$,
and second its dependence on the variable $\beta$ (or $x_{\bar{p}}$). Such a
calculation has been performed by Kaidalov et al.\ \cite{K}. In this calculation,
which we call KKMR, the hard scattering cross section for the diffractive
production of
dijets was supplemented by screening or absorptive corrections on the basis of
eikonal corrections in impact parameter ($b$) space. The parameters of the
eikonal were obtained from a two-channel description of high-energy inelastic
diffraction. The exponentiation of the eikonal stands for the exchange of
multi-Pomeron contributions, which violate Regge and QCD factorization and modify
the predictions based on single Pomeron and/or Regge exchange. The obtained
suppression factor $S$ is not universal, but depends on the details of the hard
subprocess as well as on the kinematic configurations. The first important
observation in the KKMR analysis is that in the Tevatron dijet analysis the mass
squared of the produced dijet system $M_{\rm JJ}^2=
x_p\beta \xi s$ as well as $\xi$ are almost constant, so that small $\beta$
implies large $x_p$.
The second important ingredient in the KKMR model is the assumption that the
absorption cross section of the valence and the sea components,
where the latter includes the gluon, of the incoming proton are different, in
particular, that the valence and sea components correspond to smaller and larger
absorption. For large $x_p$ or small $\beta$, the valence quark contribution
dominates, which produces smaller absorptive cross sections as compared to the
sea quark and gluon contributions, which dominate at small $x_p$. Hence the
survival probability (or suppression factor) increases as $x_p$ increases and
$\beta$ decreases. In Ref.\ \cite{K}, the convolution of the old H1 DPDFs
\cite{H1} and the $\beta$-dependent absorption corrections produced a
$F^D_{\rm JJ}(\beta)$-distribution corrected for the soft rescattering, which was
in very good agreement with the corresponding experimental distribution in the CDF
publication \cite{6} (see Fig.\ 4 in \cite{K}). We have no doubt that using our
single-diffractive NLO cross sections based on the more recent DPDFs of H1
\cite{4} will lead to a very similar result.

An alternative model for the calculation of the suppression factor was developed
by Gotsman et al.\ \cite{G}. However, these authors did not convolve their
suppression mechanism with the hard scattering cross section. Therefore a direct
comparison to the CDF data is not possible.

At variance with the above discussion of diffractive dijet production in
hadron-hadron scattering, the survival probability in diffractive dijet
photoproduction was found to be larger ($\simeq0.5$ for global suppression,
$\simeq0.3$ for resolved photon suppression only) and fairly independent of
$\beta$ (or $z_{\p}$) \cite{12,15}. This can be explained by the fact that
the HERA analyses are restricted to large values of $x_\gamma\geq 0.1$ (as opposed
to small and intermediate values of $x_p=0.02$ ... $0.2$), where direct
photons or their fluctuations into perturbative or vector meson-like valence
quarks dominate. The larger suppression factor in photoproduction corresponds
also to the smaller center-of-mass energy available at HERA.


\section{Conclusions}

In conclusion, we have performed the first next-to-leading order calculation of
single-diffractive and non-diffractive cross sections for dijet production in
proton-antiproton collisions at the Tevatron, using recently obtained parton
densities in the (anti-)proton from global fits and in the Pomeron from inclusive
deep-inelastic scattering and DIS dijet production at HERA. The normalized
distributions in the average transverse energy and rapidity of the two jets
agreed well with those measured by the CDF collaboration at two different
center-of-mass energies of $\sqrt{s}=1800$ and 630 GeV. However, the ratios
of single-diffractive and non-diffractive cross sections had two be multiplied by
factors of about 0.05 and up to 0.3, depending on the momentum fraction of the
parton in the antiproton, the center-of-mass energy, the order of the calculation,
and the DPDF. Assuming Regge factorization, the ratios of cross sections were
interpreted as ratios of effective diffractive structure functions, exhibiting
similar suppression factors.

We found that the ratios of SD over ND $K$-factors of 1.35 and 1.6 at $\sqrt{s}=
1800$ and 630 GeV, respectively, were partially compensated by the simplification
inherent in the definition of the effective structure functions, but that the
suppression factors were still smaller at NLO than at LO. They were also less
dependent on the momentum fraction of the parton in the Pomeron at NLO than at
LO, in particular at the lower center-of-mass energy and to a smaller extent also
for the more restricted kinematics at the higher $\sqrt{s}$. The DPDF fit by the
H1 collaboration using DIS dijet data to better constrain the gluon density
in the Pomeron showed the most constant behavior.

We pointed out that all suppression factors obtained so far must be corrected by
a factor of $1.23\pm 0.03(stat.)\pm 0.16(syst.)$ due to the fact that the DPDFs
were obtained from H1 data that includes diffractive dissociation, while the
CDF data were triggered on a leading antiproton. We also recalled that the
remaining momentum-fraction dependence can be explained by a two-channel
eikonal model that predicts different behaviors for the regions dominated by
valence quarks and sea quarks and gluons in the proton. This is in contrast
to the constant behavior observed in photoproduction, which is governed by
direct photon or valence-like quark contributions.
We finally confirmed that the single-diffractive data are dominated by a single
Pomeron exchange, since its momentum fraction dependence in the antiproton is well
described in shape by the Pomeron flux factors fitted to the H1 DIS data.
An additional Reggeon exchange would add only two to eight percent to the
single-diffractive cross section for smaller and larger values of the Reggeon
momentum fraction.

\acknowledgments
We thank K.\ Hatakeyama for useful discussions concerning the CDF data
analyses.
This work has been supported by the Theory-LHC-France initiative of the
CNRS/IN2P3.



\begin{thebibliography}{99}

\bibitem{1}
 P.D.B.\ Collins, {\em An Introduction to Regge Theory and High-Energy
 Physics}, Cambridge University Press, Cambridge (1977).

\bibitem{2}
 V.N.\ Gribov and L.N.\ Lipatov, Sov.\ J.\ Nucl.\ Phys.\ {\bf 5}, 438 (1972)
 and Sov.\ J.\ Nucl.\ Phys.\ {\bf 20}, 94 (1975);
 G.\ Altarelli and G.\ Parisi, Nucl.\ Phys.\ B {\bf 126}, 298 (1977);
 Y.L.\ Dokshitzer, Sov.\ Phys.\ JETP {\bf 46}, 641 (1977).

\bibitem{3}
 ZEUS Collaboration, S.\ Chekanov et al., Eur.\ Phys.\ J.\ C {\bf 38}, 43
 (2004).

\bibitem{4}
 H1 Collaboration, A.\ Aktas et al., Eur.\ Phys.\ J.\ C {\bf 48}, 715 (2006).

\bibitem{5}
 J.C.\ Collins, Phys.\ Rev.\ D {\bf 57}, 3051 (1998)
 [Erratum ibid.\ {\bf 61}, 019902 (2000)];
 J.\ Phys.\ G {\bf 28}, 1069 (2002).

\bibitem{6}
 CDF Collaboration, T.\ Affolder et al., Phys.\ Rev.\ Lett.\ {\bf 84}, 5043
 (2000).

\bibitem{7}
 H1 Collaboration, paper 980, submitted to the 31st {\em Int.\ Conf.\
 on High-Energy Physics} (ICHEP02), Amsterdam (2002).

\bibitem{8}
 M.\ Klasen and G.\ Kramer, in Proc.\ of the 12th Int. Workshop on 
 {\em Deep Inelastic Scattering } (DIS04), eds.\ D. Bruncko, I.\
 Ferencei and P.\ Strizenec, Kosize, Inst.\ Exp.\ Phys., SAS, p.\ 492
 (2004), hep-ph/0401202.

\bibitem{9}
 M.\ Klasen and G.\ Kramer, Eur.\ Phys.\ J.\ C {\bf 38}, 93 (2004).

\bibitem{10}
 H1 Collaboration, paper 987, submitted to the 31st {\em Int.\ Conf.\
 on High-Energy Physics} (ICHEP02), Amsterdam (2002);
 paper 087, submitted to the Int.\ {\em Europhysics Conf.\ on High-Energy
 Physics} (EPS03), Aachen (2003);
 F.P.\ Schilling, Eur.\ Phys.\ J.\ C {\bf 33}, S530 (2004).

\bibitem{11}
 ZEUS Collaboration, abstract 6-0249, contributed to the 32nd 
 {\em Int.\ Conf.\ on High-Energy Physics} (ICHEP04), Beijing (2004).

\bibitem{12}
 H1 Collaboration, A.\ Aktas et al., Eur.\ Phys.\ J.\ C {\bf 51}, 549
 (2007).

\bibitem{13}
 ZEUS Collaboration, S.\ Chekanov et al., Eur.\ Phys.\ J.\ C {\bf 55},
 177 (2008).

\bibitem{14}
 H1 Collaboration, K.\ Cerny et al.\, in Proc.\ of the 16th Int.\
 Workshop on {\em Deep-Inelastic Scattering} (DIS08), London (2008), 
 {\tt http://dx.doi.org/10.3360/dis.2008.69}.

\bibitem{15}
 M.\ Klasen and G.\ Kramer, Mod.\ Phys.\ Lett.\ A {\bf 23}, 1885 (2008);
 Proc.\ of the Workshop on {\em HERA and the LHC}, eds.\ H.\ Jung et al.,
 DESY-PROC-2009-02, Mar 2009, arxiv:0903.3861;
 arXiv:0808.3700;
 LPSC 08-115 (to be published);
 G.\ Kramer, Nucl.\ Phys.\ B (Proc.\ Suppl.) {\bf 191}, 231 (2009).

\bibitem{16}
 CDF Collaboration, D. Acosta et al., Phys.\ Rev.\ Lett.\ {\bf 88},
 151802 (2002).

\bibitem{17}
 M.\ Klasen, T.\ Kleinwort and G.\ Kramer, Eur.\ Phys.\ J.\ direct
 C {\bf 1}, 1(1998) and the earlier papers quoted there;
 for a review see M. Klasen, Rev.\ Mod.\ Phys.\ {\bf 74}, 1221 (2002).

\bibitem{18}
 CTEQ Collaboration, P.M.\ Nadolsky et al., Phys.\ Rev.\ D {\bf 78},
 013004 (2008).

\bibitem{19}
 CTEQ Collaboration, I.\ Pumplin et al., JHEP {\bf 07}, 021 (2002).

\bibitem{20}
 H1 Collaboration, A.\ Aktas et al., JHEP {\bf 10}, 042 (2007).

\bibitem{20a}
 K.\ Hatakeyama, private communication.

\bibitem{21}
 M.\ Klasen and G.\ Kramer, Phys.\ Lett.\ B {\bf 366}, 385 (1996); 
 S.\ Frixione and G.\ Ridolfi, Nucl.\ Phys.\ B {\bf 507}, 315 (1997).

\bibitem{22}
 See {\tt http:/physics.rockefeller.edu/hatake/phys/sdjj$\_$1800$\_$prl.html}.

\bibitem{23}
 M.\ Gl\"uck, E.\ Reya and A.\ Vogt, Eur.\ Phys.\ J.\ C {\bf 5}, 461 (1998).

\bibitem{24}
 J.\ Owens, Phys.\ Rev.\ D {\bf 30}, 943 (1984).

\bibitem{H1LP}
 H1 Collaboration, A.\ Aktas et al., Eur. Phys. J. C {\bf 48}, 749 (2006).

\bibitem{K}
 A.B.\ Kaidalov, V.A.\ Khoze, A.D.\ Martin and M.G.\ Ryskin, Eur.\
 Phys.\ J.\ C {\bf 21}, 521 (2001).

\bibitem{H1}
 H1 Collaboration, T.\ Ahmed et al., Phys.\ Lett.\ B {\bf 348}, 681 (1995);
 H1 Collaboration, C.\ Adloff et al., Z.\ Phys.\ C {\bf 76}, 613 (1997).

\bibitem{G}
 G.\ Gotsman, E.\ Levin, U.\ Maor, E.\ Naftali and A.\ Prygarin,
 Proc.\ of the Workshop on {\em HERA and the LHC}, part A, p.\ 221 (2005).

\end{thebibliography}
\end{document}